\documentclass{article}
\usepackage{url} 
\usepackage{authblk}
\usepackage{lineno}
\usepackage{amsmath}
\usepackage{amssymb}
\usepackage{mathrsfs}
\usepackage{amsmath}
\usepackage{appendix}
\usepackage{graphicx}
\usepackage{dsfont}
\usepackage{float}
\usepackage{diagbox} 
\usepackage{subfigure}

\usepackage{graphics}
\usepackage{csquotes}
\usepackage[all]{xy}

\PassOptionsToPackage{hyphens}{url}
\usepackage{tikz-cd}
\usepackage{braket}
\usepackage{xcolor}
\usepackage{verbatim}
\usepackage{float}
\usepackage{mathrsfs}
\usepackage{extarrows}
\usepackage{tikz}
\usetikzlibrary{shapes,arrows,positioning}
\usepackage{mathrsfs}
\usepackage{mathtools}

\usepackage{subfig}
\usepackage{wrapfig}
\usepackage{gensymb}
\usepackage{algorithm} 
\usepackage{algpseudocode} 
\usepackage{fancyhdr}
\usepackage[inline]{trackchanges} 
\usepackage{soul}

\newcommand{\bb}{\mathbf b}

\newcommand{\bk}{\mathbf k}

\newcommand{\bu}{\mathbf u}

\newcommand{\cS}{\mathcal S}

\newtheorem{rem}{Remark}[section]

\title{{\bf A non-column based, fully unstructured
implementation of Kessler’s microphysics with
warm rain using continuous and discontinuous
spectral elements}}

\author[1]{Yassine Tissaoui\thanks{yt277@njit.edu}}
\author[1]{Simone Marras\thanks{smarras@njit.edu}}
\author[2]{Annalisa Quaini}
\author[3]{Felipe A.\ V.\ de Brangaca Alves}
\author[3]{Francis X. Giraldo}

\affil[1]{Department of Mechanical Engineering, New Jersey Institute of Technology, Newark, New Jersey}
\affil[2]{Department of Mathematics, University of Houston, Houston, Texas}
\affil[3]{Department of Applied Mathematics, Naval Postgraduate School, Monterey, California}

\date{\today}
\begin{document}

\maketitle

\begin{abstract}
Numerical weather prediction is pushing the envelope of grid resolution at local and global scales alike. Aiming to model topography with higher precision, a handful of articles introduced unstructured vertical grids and tested them for dry atmospheres. The next step towards effective high-resolution unstructured grids for atmospheric modeling requires that also microphysics is independent of any vertical columns, in contrast to what is ubiquitous across operational and research models. In this paper, we present a non-column based continuous and discontinuous spectral element implementation of Kessler's microphysics with warm rain as a first step towards fully unstructured atmospheric models.
We test the proposed algorithm against standard three-dimensional benchmarks for precipitating clouds and show that the results are comparable with those presented in the literature across all of the tested effective resolutions. While presented for both continuous and discontinuous spectral elements in this paper, the method that we propose can very easily be adapted to any numerical method utilized in other research and legacy codes.
\end{abstract}

\section*{Plain Language Summary}
The earth climate is warming faster than ever. While climate models are the tool available to scientists to forecast its future evolution, they are biased by uncertainties that are, arguably, mostly embedded in the modeling of clouds. Thanks to the advent of exascale computing, a reduction of cloud modeling uncertainties can be expected by simulating clouds at higher and higher resolutions. While uniform high resolution across the whole domain is ideal, for computational efficiency reasons scientist are likely to increase the model resolution in some regions more than others not only in the horizontal direction ---which is a standard approach--- but also along the vertical direction. Grid refinement in the vertical direction, however, may lead to the loss of the vertical  structure of the grid columns, affecting the usability of column-based physics packages that are used to model clouds and precipitation. To overcome this problem, we present an algorithm to solve the equations that model precipitating clouds along arbitrarily shaped grids in any spatial direction. This approach is advantageous from a modeling perspective as well as from a computational one because it allows full flexibility of the domain partitioning algorithms when hundreds of thousands of parallel processors are used.

%
%

\section{Introduction}
Exascale computing on hybrid architectures is expected to become
available by the start of 2023. Massive parallelism will
enable the use of very fine grids for computational simulations. This is especially attractive
for climate and weather simulations as more physical processes will be resolved instead of parameterized. For example, the use of sufficiently refined meshes makes it possible for atmospheric models to resolve extreme precipitation events more precisely than usually aimed for nowadays \cite{Iorioetal2004,Teraietal2018,Wehneretal2014,Atlasetal2005,Caldwelletal2019,Bacmeisteretal2014}. If highly 
refined meshes for climate and weather simulations are also unstructured, it is possible to heighten the resolution of topographical features, including those that have been classically smoothed for the purpose of stabilizing global climate models \cite{Lauritzenetal2015}.
Poor topography resolution makes precise weather forecast challenging \cite{GiorgiMarinucci1996}, especially in the vicinity of steep mountain ranges \cite{yamazakiEtAl2022} such as, for example, the Himalayan region. Better resolved topography and coastal boundaries have been shown to improve the accuracy of simulations involving orographic precipitation and sea breeze effects \cite{Caldwelletal2019,DelworthetAl2012,DuffyetAl2003,PopeStratton2002,Loveetal2011}. This paper presents the first implementation of a method capable of solving the fully compressible Euler equations with moisture, cloud formation, and warm rain on three-dimensional fully unstructured grids. It aims to show that it is possible to effectively implement a traditionally column-reliant parameterization on vertically unstructured meshes.

Despite the fact that there has been interest in using unstructured grids since the 1960s \cite{Nikiforakis2009}, most of the operational and research weather forecast models 
are constrained by vertically structured and column-based grids, even in the cases when non-structured discretizations are used in the horizontal direction.
While horizontally unstructured meshes are often utilized (e.g. \cite{CESM2, dennis_2011}), vertically unstructured grids are not. This is due to the column constraints imposed by the microphysics packages that have been historically used. 

The first two atmospheric research models to adopt unstructured grids in the vertical direction were presented by \cite{aubryVazquez2010AIAAproc} and  \cite{smolarkiewiczEtAl2013}, with \cite{Szmelteretal2015} extending the latter to unstructured tetrahedral grids  in 2015. 
At the time of writing this article, the latest in this series of efforts was published by \cite{liEtAl2021}. All of them demonstrate that the use of unstructured grids combined with adaptive mesh refinement reduces the numerical errors for dry mountain waves problems with steep orography, even at high resolutions. 
Large numerical errors when using structured grids to represent steep topography are a well known problem summarized by, e.g., \cite{baldauf2021}, which shows that simulations run with the COSMO model \cite{COSMO_project} break down with slopes larger than approximately 30 degrees.
The choice of structured grids is motivated by the fact that the inclusion of microphysical processes has typically relied on a column-based, vertically structured implementation. Ever since the 1960s and 1970s when some of the first simulations of clouds and precipitation were performed utilizing microphysical parametrizations \cite{klempWilhelmson1978,kessler1969,soongOgura1973,weismanKlemp1982}, the implementation of these parameterizations has always relied on column-based grids.
Although interpolation from the native grid to a physics grid is usually required, the native grid in all of the operational and research models depends on a column-based structure. 

This paper presents a fully unstructured discretization of the compressible Euler equations with moisture to model clouds and precipitation. To support non-column based precipitation, we approximated the transport equation governing precipitation by means of the same approximation of the underlying dynamics model (i.e., the compressible Euler equations). To achieve this, we modified the Kessler's microphysics implementation in the Nonhydrostatic Unified Model of the Atmosphere (NUMA) \cite{kellyGiraldo2012}. In this way, we leverage the natural unstructured nature of the element-based Galerkin discretization \cite{giraldoBOOK} on which NUMA relies. We test the new implementation for both continuous and discontinuous elements (e.g., see \cite{abdiGiraldo2015} for how this can be achieved in the same source code). Other models that use either continuous or discontinuous spectral elements for atmospheric flows are, e.g., CESM2 \cite{CESM2}, E3SM\cite{Caldwelletal2019}, both via the CAM-SE dycore \cite{dennis_2011}, and ClimateMachine \cite{sridharEtAl2019CLIMALES}.

We show that with a simple modification of the Kessler precipitation routine, the spectral element method is capable of simulating rain precipitation through sedimentation on fully unstructured grids that do not rely on the vertical columns of a Cartesian grid. This is done in the typical spectral/finite element fashion of solving the local equations of motion on a reference element before projecting the local solution back to the physical space. This makes it possible to solve the equations of motion without any regard for the type of grid (structured or unstructured). The only constraint is that the solution quality will depend on the accuracy of the metric terms used to map the physical elements to the reference element \cite{giraldoBOOK,nelsonKoprivaCurvedGrids2016}. We test this method in 3D by performing several squall lines \cite{rotunnoKlempWeisman1988, weismanEtAl1988} and supercell \cite{skamarockEtAl2012} simulations. We show that this method is able to produce results comparable to those available in the literature. This work will help lead the way towards moist-air simulations of flow over steep orography using unstructured grids, and possibly both horizontal and vertical adaptive mesh refinement.

Finally, this approach has important consequences on the parallel efficiency for very high resolution atmospheric simulations because Message Passing Interface (MPI) is no longer limited to a column based subdivision of the domain, but will allow for a parallel load balancing decomposition in any direction. 

The remainder of the paper is organized as follows. The governing equations are presented in \S~\ref{sec:problem}. The numerical approximation of the governing equations, including the details of the discretization of the rain equation and the algorithm for non-column-based rain sedimentation are presented in \S~\ref{sec:num_meth}.
The numerical results are described in \S~\ref{sec:num_res}. The conclusions are drawn in \S~\ref{sec:conclusions}.

\section{Problem definition}\label{sec:problem}

Moist air is a mixture of dry air with density $\rho$, water vapor with density $\rho_v$, and suspended cloud condensate with density $\rho_c$. 
The mass fractions of water vapor and cloud water are defined as $q_v=\rho_v/\rho$ and $q_c=\rho_c/\rho$, 
respectively. In addition, let $\rho_r$ be the rain density and $q_r=\rho_r/\rho$ the rain mass fraction. Warm rain is assumed (No ice formation or precipitation takes place).
We denote by $c_p$ and $c_v$ the specific heat capacities at constant pressure and volume for dry air. The specific gas constants of dry air and vapor are denoted by $R_d$ and $R_v$ and set $\epsilon = \frac{R_d}{R_v}$.
Let: 
\begin{align}
\theta = (1+\epsilon q_v)  \frac{T}{\pi}, \quad\text{with } \pi = \left( \frac{p}{p_s} \right)^{\frac{R_d}{c_p}}, \label{eq:theta}
\end{align}
be the virtual potential temperature, where $T$ is the absolute temperature and $p_s = 10^5$ Pa is the ground surface pressure.
Finally, let $\bu$ be the wind velocity.

We consider a fixed spatial domain
$\Omega$ and a time interval of interest $(0,t_f]$.
Balance of mass, momentum, and potential temperature for moist air in terms of
\emph{prognostic variables} $\rho$, $\bu$, and $\theta$ in conservative form  are given by: 
\begin{align}
&\frac{\partial \rho}{\partial t} + \nabla \cdot (\rho \bu)= 0 &&\text{in } \Omega \times (0,t_f], \label{eq:mass}  \\
&\frac{\partial (\rho \bu)}{\partial t} +  \nabla \cdot \left( \rho \bu \odot \bu\right) =-\nabla p + \rho \bb &&\text{in } \Omega \times (0,t_f],  \label{eq:mom_c} \\
& \frac{\partial (\rho \theta)}{\partial t} +  \nabla \cdot \left(\rho \theta \bu\right) = \rho \cS_{\theta} &&\text{in } \Omega \times (0,t_f].
\label{eq:ent_c}
\end{align}
where $\bb$ is the total buoyancy. We have $\bb = - (1 + \epsilon q_v - q_c - q_r)g\widehat{\bk}$,  where $g=9.81~{\rm m/s^2}$ 
is the magnitude of the acceleration of gravity, and $\widehat{\bk}$ is the unit vector aligned with the vertical axis $z$.
Finally, the source/sink term $\cS_{\theta}$ in \eqref{eq:ent_c} describes latent heat release–uptake during phase changes of moisture variables and is detailed in Sec.~\ref{sec:micro}. Eq.~\eqref{eq:mom_c} and \eqref{eq:ent_c} can be rewritten in non-conservative form as follows:
\begin{align}
& \frac{\partial \bu}{\partial t} + \bu \cdot \nabla \bu = - \frac{1}{\rho} \nabla p  + \bb &&\text{in } \Omega \times (0,t_f],  \label{eq:mom} \\
& \frac{\partial \theta}{\partial t} + \bu \cdot \nabla \theta = \cS_{\theta} &&\text{in } \Omega \times (0,t_f].
\label{eq:ent}
\end{align}

A thermodynamics equation of state for the pressure  of moist air $p$ is needed for closure.
We assume that $p$ is the sum of the 
partial pressures of dry air and vapor ($p_d$ and $p_v$, respectively), both taken to be ideal gases. Thus, neglecting the volume of the condensed phase, the equation of state relating $p$ to $\rho$ and $T$ is given by:
\begin{align}
p = p_d + p_v = \rho R_d T + \rho q_v R_v T = \rho R_d T (1+\epsilon q_v). \label{eq:p}
\end{align}

To facilitate the numerical solution of system \eqref{eq:mass}-\eqref{eq:ent_c} or \eqref{eq:mass}, \eqref{eq:mom}-\eqref{eq:ent}, we write density, pressure, 
and potential temperature as the sum of
their mean hydrostatic values and fluctuations:
\begin{align}
\rho(x,y,z,t) &= \rho_0(z) + \rho'(x,y,z,t), \label{eq:rho_split} \\
\theta(x,y,z,t) &= \theta_0(z) + \theta'(x,y,z,t), \label{eq:theta_split} \\
p(x,y,z,t) &= p_0(z) + p'(x,y,z,t). \label{eq:p_split}
\end{align}
Note that the hydrostatic reference states are functions
of the vertical coordinate $z$ only. Hydrostatic balance relates
$p_0$ to $\rho_0$ as follows:
\begin{align}
\frac{dp_0}{dz} = - \rho_0 g. \label{eq:hydro_bal}
\end{align}
Plugging \eqref{eq:rho_split}-\eqref{eq:p_split} into 
\eqref{eq:mass}-\eqref{eq:ent_c} and accounting for \eqref{eq:hydro_bal} leads to:
\begin{align}
& \frac{\partial \rho'}{\partial t}  + \nabla \cdot ((\rho_0+ \rho') \bu)  = 0, \label{eq:mass_c_split} \\
&\frac{\partial ((\rho_0+ \rho') \bu)}{\partial t} +  \nabla \cdot \left( (\rho_0+ \rho') \bu \otimes \bu \right)  + \rho' g \widehat{\bk} = - \nabla p' + (\rho_0+ \rho') \widetilde{\bb},  \label{eq:mom_c_split} \\
&\frac{\partial ((\rho_0+ \rho') (\theta_0 + \theta' ))}{\partial t} +  \nabla \cdot \left((\rho_0+ \rho') \theta' \bu \right) +  \nabla \cdot \left((\rho_0+ \rho') \theta_0 \bu \right) = (\rho_0+ \rho') \cS_{\theta}, \label{eq:ent_c_split}
\end{align}
where $\widetilde{\bb} = -\left( \frac{\rho'}{\rho_0 + \rho'} + \epsilon q_v - q_c - q_r\right)g\widehat{\bk}$ is a modified total buoyancy.
Following a similar procedure for Eq.~\eqref{eq:mom}-\eqref{eq:ent}, we obtain
\begin{align}
& \frac{\partial \rho'}{\partial t}  + \nabla \cdot ((\rho_0+ \rho') \bu) = 0, \label{eq:mass_split} \\
& \frac{\partial \bu}{\partial t} + \bu \cdot \nabla \bu = - \frac{1}{\rho_0 + \rho'} \nabla p'  + \widetilde{\bb},  \label{eq:mom_split} \\
& \frac{\partial \theta'}{\partial t} + \bu \cdot \nabla \theta_0 + \bu \cdot \nabla \theta' = \cS_{\theta}.
\label{eq:ent_split}
\end{align}

\begin{rem}\label{rem1}
To preserve numerical stability of the solution, we add an  artificial diffusion term with a constant diffusivity coefficient $\beta$ to equation sets  \eqref{eq:mass_c_split}-\eqref{eq:ent_c_split} and \eqref{eq:mass_split}-\eqref{eq:ent_split}; the units of $\beta$ are given consistently with the equations at hand. The term 
$\beta\nabla^2 \bu$ is added to the right-hand side of the momentum equation, while the term  
$\beta\nabla^2 \theta' $ is added to the right-hand side of the equation of the potential temperature.
\end{rem}

\begin{rem}\label{rem2}
While we usually stabilize NUMA simulations by leveraging the eddy viscosity from an LES model (see \cite{marrasNazarovGiraldo2015, reddyEtAl2021}), in this paper we consider artificial viscosity with constant $\beta$ as it is done in \cite{gabersekGiraldoDoyle2012, skamarockEtAl2012} whose results we are testing against.
\end{rem}

Next, we write the balance equations for $q_v$ and $q_c$ in conservative form: 
\begin{align}
 &\frac{\partial (\rho q_v)}{\partial t} + \nabla \cdot (\rho q_v \bu) =  \rho \cS_{v}&&\text{in } \Omega \times (0,t_f], \label{eq:moist1_0} \\
 &\frac{\partial (\rho q_c)}{\partial t} + \nabla \cdot (\rho q_c \bu) = \rho \cS_{c}&&\text{in } \Omega \times (0,t_f], \label{eq:moist2_0}
 \end{align}
and non-conservative form: 
\begin{align}
\frac{\partial q_v}{\partial t} + \bu \cdot \nabla q_v &=  \cS_{v} &&\text{in } \Omega \times (0,t_f], \label{eq:moist1_1} \\
\frac{\partial q_c}{\partial t} + \bu \cdot \nabla q_c &= \cS_{c} &&\text{in } \Omega \times (0,t_f]. \label{eq:moist2_1}
\end{align}

The source/sink terms on the right-hand side in the equations above are related to conversion rates. In particular, we have:
\begin{align}
\cS_{v} = C(q_c \rightarrow q_v) + C(q_r \rightarrow q_v), \quad
\cS_{c} = C(q_v \rightarrow q_c) + C(q_r \rightarrow q_c), \quad \cS_t = \cS_{v} + \cS_{c},
\end{align}
where the terms $C(q_\phi \rightarrow q_\psi) = - C(q_\psi \rightarrow q_\phi)$ represent the conversion of species $\phi$ to species $\psi$. All of these terms, which account for processes such as evaporation of cloud condensate, are provided by the microphysics equations reported in Sec.~\ref{sec:micro}.

Precipitating water (rain) is treated in the same manner. Letting $w_{r}$ be the fall speed of rain (provided by the microphysics equations), we can write the conservation law for rain in conservative form:
\begin{align}
 \frac{\partial (\rho q_r)}{\partial t} + \nabla \cdot (\rho q_r (\bu - w_r \widehat{\bk})) =  \rho \cS_r \quad \text{in } \Omega \times (0,t_f], \label{eq:qr_0}
\end{align}
and non-conservative form:
\begin{align}
\frac{\partial q_r}{\partial t} + \bu \cdot \nabla q_r =  \cS_r + \frac{1}{\rho} \frac{\partial}{\partial z} \left(\rho q_r w_r\right)  \quad \text{in } \Omega \times (0,t_f], \label{eq:qr_1}
\end{align}
with 
\begin{align}
\cS_r = C(q_v \rightarrow q_r) + C(q_c \rightarrow q_r).
\end{align}

In summary, the conservative form of the
atmospheric model considered in this paper is given by \eqref{eq:mass_c_split}-\eqref{eq:ent_c_split}, \eqref{eq:moist1_0}-\eqref{eq:moist2_0}, \eqref{eq:qr_0} and \eqref{eq:p}, while its non-conservative form
is given by \eqref{eq:mass_split}-\eqref{eq:ent_split}, \eqref{eq:moist1_1}-\eqref{eq:moist2_1}, \eqref{eq:qr_1} and \eqref{eq:p}. In both cases, the problem has to be supplemented with proper initial and boundary conditions that will be specified in Sec.~\ref{sec:num_res}. 


\subsection{Microphysical parameterization}\label{sec:micro}


The terms on the right-hand sides of Eq.~\eqref{eq:ent_c_split}, \eqref{eq:moist1_0}, \eqref{eq:moist2_0}, and \eqref{eq:qr_0}, and their respective non-conservative counterparts are
defined according to \cite{klempWilhelmson1978}. Let $q_{vs}$ be the saturation water vapor fraction. To determine $q_{vs}$ we use Teten's formula following \cite{klempWilhelmson1978}. The evaporation of cloud water is given by:
\begin{align}
C(q_c \rightarrow q_v) = -  C(q_v \rightarrow q_c)  = \frac{\partial q_{vs}}{\partial t}.
\end{align}
This is computed with the saturation adjustment approach of Soong and Ogura \cite{soongOgura1973}. The evaporation of rain, i.e.~conversion rate $C(q_r \rightarrow q_v) = -  C(q_v \rightarrow q_r)$, is taken directly from \cite{klempWilhelmson1978}, which uses an approach similar to \cite{oguratakahashi}. We have 
\begin{equation}\label{raineq}
    C(q_c \rightarrow q_r) = -C(q_r \rightarrow q_c) = A_r + C_r,
\end{equation}
where  $A_r$ and $C_r$ represent rain auto-conversion and rain accretion \cite{kessler1969}, respectively. Finally, the source/sink term in Eq.~\eqref{eq:ent_split} is given by:
\begin{equation}\label{tempeq}
    S_{\theta} = -\gamma\left(\frac{\partial q_{vs}}{\partial t} + C(q_r \rightarrow q_v) \right), \quad \gamma = \frac{L}{c_p \pi},
\end{equation}
where $L$ is the latent heat of vaporization and $\pi$ is the Exner pressure defined in \eqref{eq:theta}.

Finally, we define the terminal velocity of rain following \cite{soongOgura1973,kessler1969,klempWilhelmson1978}:
\begin{equation}\label{terminal_velocity}
    w_r = 3634(\rho q_r ^{0.1346})\left(\frac{\rho}{\rho_g}\right)^{-\frac{1}{2}},
\end{equation}
where $\rho_g$ is the reference density at the surface.
\section{Numerical method}\label{sec:num_meth}
\subsection{The Galerkin spectral element method}

In time, the equations are advanced using an implicit-explicit order 3 additive Runge-Kutta (ARK3) scheme \cite{kennedy:2003} whereby the non-linear terms of the governing equations are treated explicitly and the linear terms are treated implicitly (see \cite{giraldoEtAl2013}). As for the space discretization, we use
spectral elements and show results for both continuous and discontinuous approximations. 
This section focuses on the space discretization alone. 

To make the description of the numerical method easy to follow, we consider a generic equation of the form:
\begin{equation}\label{eq:abstraction_law}
    \frac{\partial f}{\partial t} + G(f)=0, 
\end{equation}
where $f$ is the unknown variable and $G$ is a linear functional that may contain first and second derivatives of $f$. If the equations to be solved are written in conservation form, then $G$ is the divergence of a flux. Notice that all the equations in Sec.~\ref{sec:problem} can be rewritten as
\eqref{eq:abstraction_law}.

\begin{figure}[h!]
    \centering
	\includegraphics[width=0.3\textwidth]{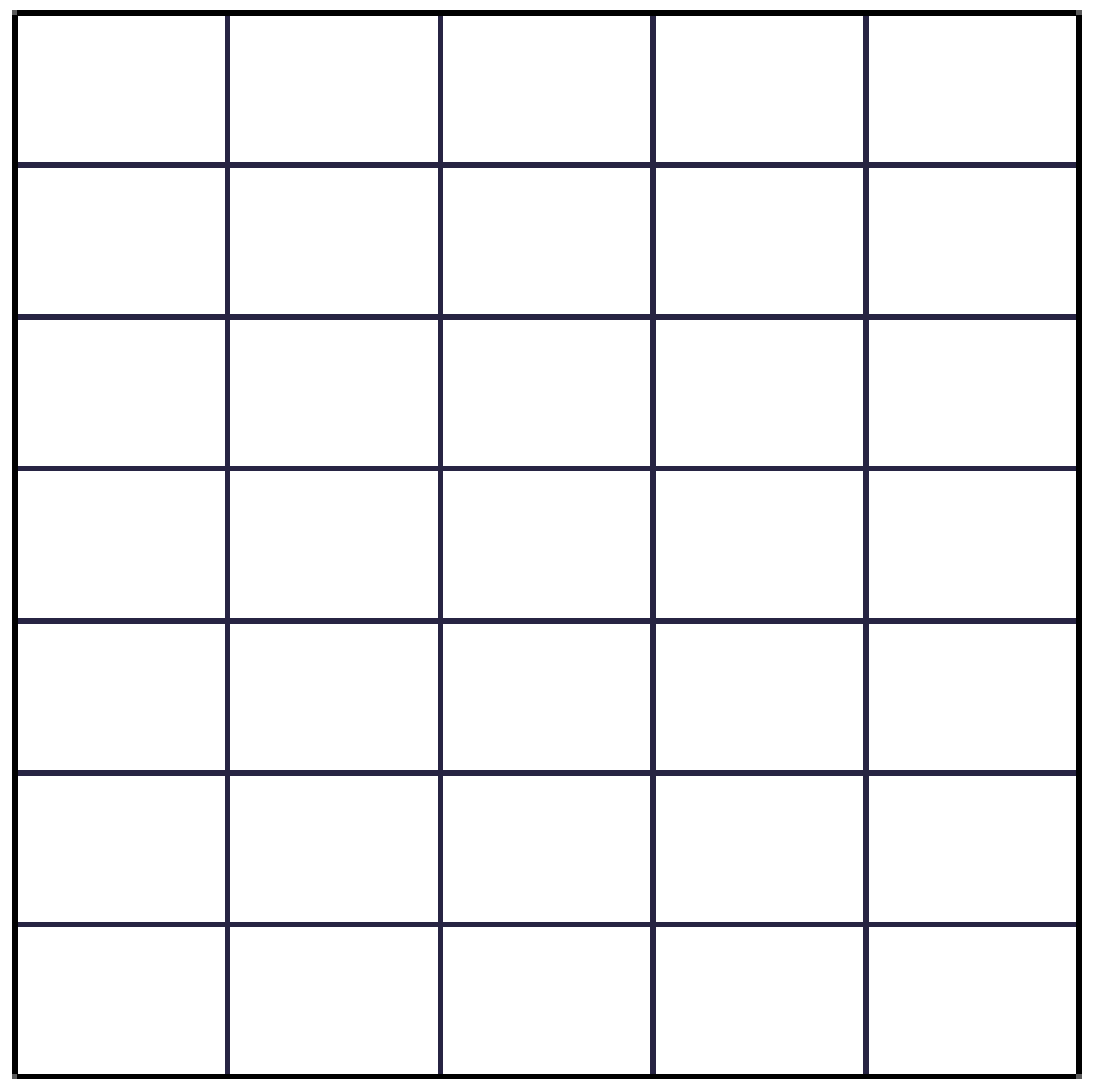}
	\includegraphics[width=0.298\textwidth]{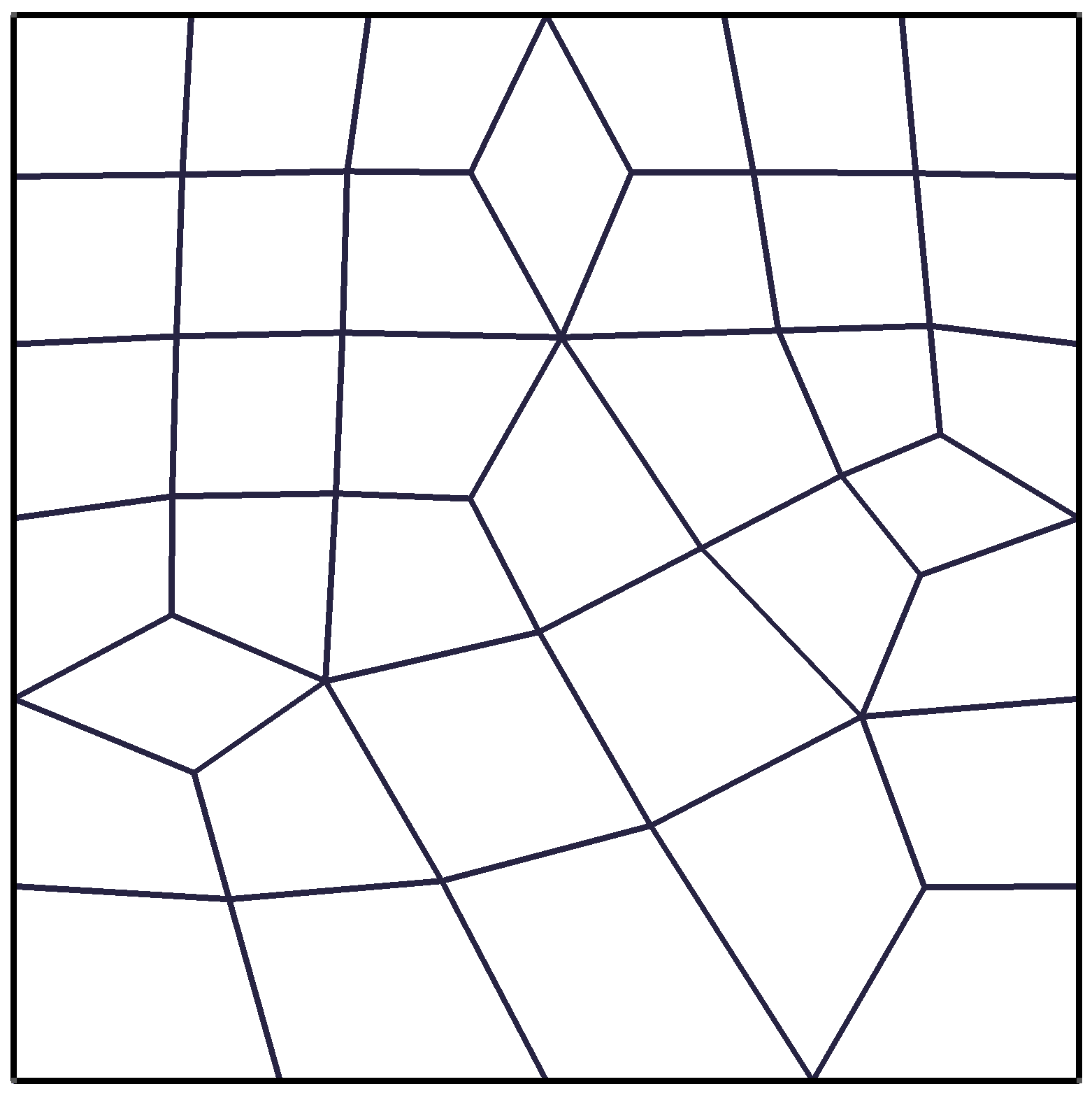}
      \caption{Examples of a structured (left) and an unstructured grid (right) made of quadrilateral elements.}
      \label{fig:Grids}
\end{figure}

We subdivide the domain $\Omega$ into a set of conforming \footnote{The condition of conformity is not strictly necessary, although it simplifies the discussion of the method. For results with non-conforming grids, the reader is referred to, e.g., \cite{koperaGiraldo2013a}.} $N_e$ hexahedral elements $\Omega_e$ of arbitrary orientation to create the discrete domain $\Omega^h$ as
\begin{equation}
    \Omega \approx \Omega^h = \bigcup_{e=1}^{N_e} \Omega_e.
\end{equation}
Fig.~\ref{fig:Grids} shows examples of a structured and unstructured grid in 2D. Using a fully unstructured grid means that structures such as the rows or columns that are seen on the left side of Fig.~\ref{fig:Grids} are no longer present.
Let $\Omega_{ref}$ be reference element:
$(\xi, \eta) \in [-1, 1]^2$ in 2D and $(\xi, \eta, \zeta) \in [-1, 1]^3$ in 3D. Regardless of whether the mesh is 
structured or unstructured, we introduce a mapping from a generic element in the  global system of coordinates, i.e.~$(x, y)$ in 2D and $(x, y,z)$ in 3D, to the reference element. Let $\mathbf{J}$ be the Jacobian matrix of this mapping.


Let $h_i$, $i=1, \dots, N+1$, be the Lagrange polynomials of degree $N$:
\begin{equation*}
     h_i(\xi) = \frac{1}{N(N+1)}\frac{(1-\xi^2)P^{'}_N(\xi)}{(\xi - \xi_i)P_N(\xi)},
\end{equation*}
where $P_N$ is the Legendre polynomial of order $N$, and $P_N^{'}$ its derivative evaluated at the point $\xi$.
The polynomials in multiple dimensions are built via a tensor product of the 1D bases, as shown below.
The remainder of this section is written for a 3D case.

For every element, we seek an approximation $f^h$ of variable $f$ of the form:
\begin{equation}\label{eq:f_h}
    f^h(\boldsymbol{\xi},t) = \sum_{l=1}^{(N+1)^3} \psi_l(\boldsymbol{\xi})\hat{f_l}(t),
\end{equation}
where $\boldsymbol{\xi}=(\xi, \eta,\zeta)$, $\hat{f_l}$ are the expansion coefficients, and $\psi_l$ are nodal basis functions defined as tensor products of the Lagrange polynomials
\begin{equation}
    \psi_l = h_i[\xi(\mathbf{x})] \odot h_j[\eta(\mathbf{x})] \odot h_k[\zeta(\mathbf{x})], \hspace{5pt} \hspace{3pt} l=i+1 + j(N+1) + k(N+1)(N+1),
\end{equation}
where $\mathbf{x}=(x,y,z)$. The Legendre-Gauss-Lobatto (LGL) points are not equidistant and represent the solutions of the following equation:
\begin{equation*}
    (1-\xi^2)P^{'}_N(\xi)=0.
\end{equation*}
The LGL points are associated with the following quadrature weights:
\begin{equation*}
    \omega(\xi_i)=\frac{2}{N(N+1)}\left[\frac{1}{P_N(\xi_i)}\right]^2
\end{equation*}
used to approximate the integrals with a Gauss quadrature rule of accuracy $\mathcal{O}(2N-1)$. Over a generic element $\Omega_e$, this is done as follows:
\begin{equation}\label{eq:Gauss_quadrature}
    \int_{\Omega_e}f(\mathbf{x})d\mathbf{x} = \int_{\Omega_{ref}}f(\boldsymbol{\xi})|\mathbf{J}(\boldsymbol{\xi})|d\boldsymbol{\xi}\approx \sum_{i,j,k=1}^{N+1}\omega(\xi_i) \omega(\eta_j) \omega(\zeta_k) f(\xi_i,\eta_j,\zeta_k)|\mathbf{J}(\xi_i,\eta_j,\zeta_k)|,
\end{equation}
where $|\mathbf{J}|$ is the determinant of the Jacobian matrix.

To approximate the solution of Eq.~\eqref{eq:abstraction_law}, let $(\cdot,\cdot)$ be the Legendre inner product on a given element $\Omega_e$:
\begin{equation*}
    (f,g)_e=\int_{\Omega_e} f({\mathbf{x}}) g({\mathbf{x}})d\mathbf{x}.
\end{equation*}
If in \eqref{eq:abstraction_law} we replace $f$ with $f^h$ as defined in \eqref{eq:f_h}, we will obtain the following residual:
\begin{equation}
    R = \frac{\partial f^h}{\partial t} + G(f^h),
\end{equation}
which is orthogonal to the expansion functions in Galerkin methods, i.e.:
\begin{equation}\label{eq:R_orth}
    (R,\psi_k)_e=0, \hspace{5pt} k=1,\dots,(N+1)^3.
\end{equation}
Taking \eqref{eq:R_orth} into account, we can now write an approximation of Eq.~(\ref{eq:abstraction_law}) on each element $\Omega_e$ as follows :
\begin{equation}\label{eq:Abstraction 2}
    \int_{\Omega_e}\psi_i(\mathbf{x}) \frac{\partial f^h(\mathbf{x},t)}{\partial t} d\mathbf{x} = - \int_{\Omega_e} \psi_i(\mathbf{x}) G(f^h(\mathbf{x},t)) d \mathbf{x}, \hspace{5pt} i=1,\dots,(N+1)^3.
\end{equation}

Let us first consider the case where $G(f)=\boldsymbol{\nabla} \cdot \mathbf{f}$, where $\boldsymbol{\nabla}= \left( \frac{\partial }{\partial x},\frac{\partial}{\partial y},\frac{\partial}{\partial z}\right)$ and $\mathbf{f} = (f,f,f)$. 

We can use the polynomial expansion to write \eqref{eq:Abstraction 2} as follows:
\begin{equation}\label{eq:first_derivative1}
    \int_{\Omega_e} \psi_i(\mathbf{x}) \sum_{j=1}^{(N+1)^3}\psi_j(\mathbf{x})\frac{\partial \hat{f}^e_j(t)}{\partial t} d \mathbf{x} = -\int_{\Omega e} \psi_i(\mathbf{x}) \sum_{j=1}^{(N+1)^3} \boldsymbol{\nabla} \psi_j(\mathbf{x})\cdot \mathbf{\hat{f}^e_j(t)} d \mathbf{x}, \hspace{5pt}, i=1,\dots,(N+1)^3,
\end{equation}
where the superscript $e$ is used to denote that the expansion is defined on an element basis and $\mathbf{\hat{f}^e_j(t)} = (\hat{f}^e_j(t),\hat{f}^e_j(t),\hat{f}^e_j(t))$. 
We can now write the mass matrix $\mathbf{M}_{ij}^e$ and the differentiation matrix $\mathbf{D}_{ij}^e$ on each element: 
\begin{align}
    \mathbf{M}_{ij}^e&=\int_{\Omega_e}\psi_i(\mathbf{x}) \psi_j(\mathbf{x}) d \mathbf{x} = \int_{\Omega_{ref}} \psi_i(\boldsymbol{\xi})\psi_j(\boldsymbol{\xi})| \mathbf{J}(\boldsymbol{\xi})| d \boldsymbol{\xi}, \label{eq:Mass} \\
    \mathbf{D}_{ij}^e&=\int_{\Omega_e}\psi_i(\mathbf{x}) \boldsymbol{\nabla}\psi_j(\mathbf{x}) d \mathbf{x} = \int_{\Omega_{ref}} \psi_i(\boldsymbol{\xi})\left(\boldsymbol{\nabla}_{\xi} \psi_j(\boldsymbol{\xi})\mathbf{J}^{-1}(\boldsymbol{\xi})\right) |\mathbf{J}(\boldsymbol{\xi})|   d \boldsymbol{\xi}, \label{eq:Differentiation}
\end{align}
with $i,j=1,\dots,(N+1)^3$ and  $\boldsymbol{\nabla}_{\xi}=\left( \frac{\partial}{\partial \xi},\frac{\partial}{\partial \eta},\frac{\partial}{\partial \zeta}\right)$.
By approximating the integrals with a quadrature rule, we obtain:
\begin{align}
    \mathbf{M}^e_{ij}&=\sum_{k=1}^{N+1} \sum_{m=1}^{N+1}\sum_{n=1}^{N+1}\omega(\xi_k,\eta_m,\zeta_n) 
    \psi_i(\xi_k,\eta_m,\zeta_n) \psi_j(\xi_k,\eta_m,\zeta_n)
    |\mathbf{J}(\xi_k,\eta_m,\zeta_n)|, \label{eq:M} \\
    \mathbf{D}^e_{ij}&=\sum_{k=1}^{N+1} \sum_{m=1}^{N+1}\sum_{n=1}^{N+1}\omega(\xi_k,\eta_m,\zeta_n) \psi_i(\xi_k,\eta_m,\zeta_n) \boldsymbol{\nabla} \psi_j(\xi_k,\eta_m,\zeta_n)
    |\mathbf{J}(\xi_k,\eta_m,\zeta_n)|. \label{eq:D}
\end{align}

Note that $\boldsymbol{\nabla}\psi_j(\xi_k,\eta_m,\zeta_n) = \boldsymbol{\nabla}_{\xi} \psi_j(\xi_k,\eta_m,\zeta_n) \mathbf{J}^{-1}(\xi_k,\eta_m,\zeta_n)$. Then, 
the matrix form of Eq.~\eqref{eq:first_derivative1} is:
\begin{equation}\label{eq:matrix form 1}
    \mathbf{M}_{ij}^e\frac{\partial \hat{f}_j^e(t)}{\partial t} = -\mathbf{D}_{ij}\hat{f}_j^e(t), \hspace{5pt} i,j=1,\dots,(N+1)^3.
\end{equation}

Let us now consider $G(f)=\boldsymbol{\nabla}\cdot \mathbf{f} - \boldsymbol{\nabla}^2f$ in Eq.~\eqref{eq:abstraction_law}, where $\boldsymbol{\nabla}^2=\boldsymbol{\nabla} \cdot \boldsymbol{\nabla}$.
In this case, Eq.~\eqref{eq:Abstraction 2}
becomes: 
\begin{align}
\nonumber \int_{\Omega_e} \psi_i(\mathbf{x}) \sum_{j=1}^{(N+1)^3}\psi_j(\mathbf{x})\frac{\partial \hat{f}^e_j(t)}{\partial t} d \mathbf{x} =& -\int_{\Omega e} \psi_i(\mathbf{x}) \sum_{j=1}^{(N+1)^3}\boldsymbol{\nabla}\psi_j(\mathbf{x})\cdot \mathbf{\hat{f}^e_j(t)} d \mathbf{x} \\ &+ \int_{\Omega_e} \psi_i \boldsymbol{\nabla} \cdot \left[ \sum_{j=1}^{(N+1)^3} \boldsymbol{\nabla}\psi_j(\mathbf{x})\hat{f}^e_j(t)\right] d \mathbf{x}, \label{eq:46}
\end{align}
where $i,j=1,\dots,(N+1)^3$. 
After integrating by parts the second term on the right-hand side, we can rewrite \eqref{eq:46} as:
\begin{align}\label{eq:second_derivative}
    \nonumber \mathbf{M}_{ij}^e\frac{\partial \hat{f}^e_j(t)}{\partial t} =&-\mathbf{D}_{ij}^e\hat{f}^e_j(t) + \left[\psi_i(\mathbf{x})\sum_{j=1}^{N+1}\boldsymbol{\nabla}\cdot\psi_j(\mathbf{x})\hat{f}_j^e(t)\right]_{\Gamma_e} \\ &- \int_{\Omega_e}\boldsymbol{\nabla}\psi_i(\mathbf{x}) \cdot \sum_{j=1}^{N+1}\boldsymbol{\nabla}\psi_j(\mathbf{x}) \hat{f}_j^e(t) d \Omega_e  \hspace{5pt} i,j=1,\dots,(N+1)^3,
\end{align}
where $\Gamma_e$ represents the element boundary. For the sake of brevity, we assume  that the boundary term, i.e., the second term on the right-hand side in \eqref{eq:second_derivative}, 
vanishes at all element boundaries. We refer the reader to, e.g., \cite{giraldoBOOK,kellyGiraldo2012} for a detailed explanation of how this term is handled when it is not zero, as is the case for DG.
Under the assumption of vanishing boundary terms, Eq.~\eqref{eq:second_derivative}
becomes:
\begin{equation}\label{eq:second derivative lap 1}
    \mathbf{M}_{ij}^e\frac{\partial \hat{f}^e_j(t)}{\partial t} = -\mathbf{D}_{ij}^e\hat{f}^e_j(t) - \int_{\Omega_e} \boldsymbol{\nabla} \psi_i(\mathbf{x}) \cdot \boldsymbol{\nabla}\psi_j(\mathbf{x}) d \mathbf{x} \hat{f}^e_j, \hspace{5pt} i,j=1,\dots,(N+1)^3.
\end{equation}
We define the Laplacian matrix as follows:
\begin{equation}\label{eq:L0}
    \mathbf{L}^e_{ij} =\int_{\Omega_e} \boldsymbol{\nabla} \psi_i(\mathbf{x}) \cdot \boldsymbol{\nabla}\psi_j(\mathbf{x}) d \mathbf{x}= \int_{\Omega_{ref}} (\boldsymbol{\nabla}_{\xi} \psi_i(\boldsymbol{\xi}) \mathbf{J}^{-1}(\boldsymbol{\xi}))\cdot
    (\boldsymbol{\nabla}_{\xi} \psi_j(\boldsymbol{\xi}) \mathbf{J}^{-1}(\boldsymbol{\xi}))|\mathbf{J}(\boldsymbol{\xi})|d\boldsymbol{\xi}, 
\end{equation}
where $i,j=1,\dots,N+1$. 
By approximating the integral in \eqref{eq:L0} with a quadrature rule, we obtain:
\begin{equation}\label{eq:L}
    \mathbf{L}^e_{ij} = \sum_{k=1}^{N+1} \sum_{m=1}^{N+1}\sum_{n=1}^{N+1}\omega(\xi_k,\eta_m,\zeta_n)  \boldsymbol{\nabla}\psi_i(\xi_k,\eta_m,\zeta_n) \cdot \boldsymbol{\nabla}\psi_j(\xi_k,\eta_m,\zeta_n) |\mathbf{J}(\xi_k,\eta_m,\zeta_n)|,
\end{equation}
where $i,j=1,\dots,(N+1)^3$. Then, we write \eqref{eq:second derivative lap 1} as:

\begin{equation}\label{AD}
    \mathbf{M}_{ij}^e\frac{\partial \hat{f}^e_j(t)}{\partial t} = -\mathbf{D}_{ij}^e\hat{f}^e_j(t) - \mathbf{L}^e_{ij}\hat{f}^e_j(t) \hspace{5pt}, i,j=1,\dots,(N+1)^3.
\end{equation}

Next, we present briefly how the global solution is calculated depending on the choice of continuous Galerkin (CG) or discontinuous Galerkin (DG) spectral elements. The reader interested in more details on Galerkin spectral element methods is referred to, e.g., \cite{giraldoBOOK,hesthavenWarburton2008,KoprivaBook,Sherwin-KarniadakisBook}.

\paragraph{CG approximation:}
Let $\mathbf{M}$, $\mathbf{D}$, and $\mathbf{L}$ be the global mass matrix, global differentiation matrix, and global Laplacian matrix. These matrices are, in principle, assembled using Direct Stiffness Summation (DSS):

\begin{equation*}\label{eq:element law}
    \mathbf{M} = \sum_{e=1}^{N_e} \mathbf{M}^e, \hspace{5pt} \mathbf{D} = \sum_{e=1}^{N_e} \mathbf{D}^e, \hspace{5pt} \mathbf{L}=\sum_{e=1}^{N_e} \mathbf{L}^e
\end{equation*}
where $\mathbf{M}^e$ is the element mass matrix \eqref{eq:M}, $\mathbf{D}^e$ is the element differentiation matrix \eqref{eq:D}, and $\mathbf{L}^e$ is the element weak Laplacian matrix \eqref{eq:L}.
Since the same set of LGL points are used for both interpolation and integration, the global mass matrix $\mathbf{M}$ is diagonal and thus easy to invert. This is only the case if we integrate using $N+1$ LGL points as shown in \eqref{eq:Gauss_quadrature}. This type is known as inexact numerical integration, since 
the number of LGL quadrature points necessary to integrate a polynomial of order $2N$ (such as is the case for the mass matrix) up to machine precision is $N+2$. We choose to sacrifice accuracy in favor of obtaining an easily invertible mass matrix, which allows us to save considerable computational time. Additionally, it has been shown that when using polynomials of order $N\geq 4$ this type of integration has a minimal impact on accuracy, with the impact decreasing as the polynomial order is increased \cite{giraldoBOOK}. 
For the results in Sec.~\ref{sec:num_res}, we use 
$N=4$. It should be noted, however, that no global matrix is actually constructed (except for the diagonal mass matrix); the differentiation and Laplacian global matrices are never stored, only the action of these matrices on the solution vector is computed (see, e.g., \cite{giraldoBOOK}).

The global form associated with Eq.~\eqref{eq:abstraction_law} for  $G(f)=\boldsymbol{\nabla}f + \boldsymbol{\nabla}^2f$ can be written as:
\begin{equation}\label{solve}
    \frac{\partial \mathbf{f}^h}{\partial t} + \mathbf{M}^{-1}(\mathbf{D} \mathbf{f}^h + \mathbf{L}\mathbf{f}^h)=0,
\end{equation}
where $\mathbf{f}^h$ is the vector containing the nodal values of $f^h$.

\paragraph{DG approximation:}

For this kind of approximation, the global matrices are not constructed since an element communicates only with the neighboring elements through inter-element numerical fluxes. 
Thus, we write a local approximation  of Eq.~\eqref{eq:abstraction_law}, instead of a global one as in \eqref{solve}.

Let us apply integration by parts to the entries of the differentiation matrix:
\begin{equation}\label{eq:53}
\mathbf{D}^e_{ij} = \int_{\Omega_e}\psi_i(\mathbf{x}) \boldsymbol{\nabla}\psi_j(\mathbf{x}) d\mathbf{x} = \int_{\Gamma_e}\psi_i(\mathbf{x})\psi_j(\mathbf{x})\mathbf{n}^{(F,e)} d \Omega_e - \int_{\Omega_e}\boldsymbol{\nabla}\cdot\psi_i(\mathbf{x}) \psi_j(\mathbf{x}) d\mathbf{x},
\end{equation}
where $i,j=1,\dots,(N+1)^3$, $\mathbf{n}^{(F,e)}$ is the outwards facing normal of inter-element  face $F$ of the element $e$. 
The first term of the right-hand side in \eqref{eq:53}
represents an inter-element flux or a boundary flux if the element is a boundary element and it enforces the continuity of the global solution.
Notice that in a CG discretization this term vanishes as continuity is enforced via DSS.
We define the corresponding matrix as follows:
\begin{equation}\label{eq:solve_DG}
    \mathbf{F}^e_{ij} = \int_{\Gamma_e} \psi_i(\mathbf{x})\psi_j(\mathbf{x}) \mathbf{n}^{(F,e)} d\mathbf{x} \approx \sum_{F=1}^{N_F}\sum_{k=1}^{N+1}\sum_{m=1}^{N+1}\omega(\boldsymbol{\xi}_{F,km})\psi_i(\boldsymbol{\xi}_{F,km})\psi_j(\boldsymbol{\xi}_{F,km})|\mathbf{J}(\boldsymbol{\xi}_{F,km})|\mathbf{n}^{(F,e)},
\end{equation}
where $i,j=1,\dots,(N+1)^3$, $N_F$ is the number of faces for element $e$ and $\boldsymbol{\xi}_{F,km}$ denotes an integration point on the face $F$ of the element. 
The second term on the right-hand side in \eqref{eq:53} is called the weak differentiation matrix and is approximated  as follows:
\begin{align}
   \hat{\mathbf{D}}^e_{ij} &=  \int_{\Omega_e}\boldsymbol{\nabla}\psi_i(\mathbf{x}) \psi_j(\mathbf{x}) d\mathbf{x} 
   &\approx
   \sum_{k=1}^{N+1} \sum_{m=1}^{N+1}\sum_{n=1}^{N+1} \omega(\xi_k,\eta_m,\zeta_n) |\mathbf{J}(\xi_k,\eta_m,\zeta_n)| \boldsymbol{\nabla} \psi_i(\xi_k,\eta_m,\zeta_n) \psi_j(\xi_k,\eta_m,\zeta_n),\label{eq:D_weak}
\end{align}
where $i,j=1,\dots,(N+1)^3$. 

We can now rewrite \eqref{AD} for a DG discretization taking $G(f)=\boldsymbol{\nabla}\cdot f + \boldsymbol{\nabla}^2f$, which holds on each element as follows:
\begin{equation*}
     {\bf M}_{ij}^e \frac{\partial \hat{f}^e_j(t)}{\partial t} =  -\hat{\mathbf{D}}_{ij}^{e}\hat{f}_j^e(t) + \mathbf{F}_{ij}^{e}\mathbf{f}^*_j(t) - \mathbf{L}_{ij}^e \hat{f}_j^e(t)=0, \quad i,j=1,\dots,(N+1)^3,
\end{equation*}
where $\mathbf{f}^*$ represents the inter-element interface values of $\hat{f}^e_j$. We define $\mathbf{f}^*$ as follows: 
\begin{equation*}
    \mathbf{f}_j^* = \mathbf{C}(\hat{f}^e_j)-\mathbf{P}(\hat{f}^e_j)
\end{equation*}
where $\mathbf{P}$ is a penalty term and the central term $\mathbf{C}$ is defined as follows:
\begin{equation*}
    \mathbf{C}(\hat{f}^e_j) = (g(\hat{f}^{e,R}_j) + g(\hat{f}^{e,L}_j)) / 2,
\end{equation*}
where L and R refer to the left and right sides of a given inter-element interface. The function $g$ is dependent on the first derivative component of $G$ in \eqref{eq:abstraction_law} where, in this case, $G(f)=\boldsymbol{\nabla} \cdot \mathbf{f} + \boldsymbol{\nabla}^2 f$ and $g(f) = \mathbf{f}$.
The definition of $P$ depends on the choice of numerical flux. The simplest and most commonly used flux for DG is the Rusanov flux \cite{giraldoBOOK}, which gives:
\begin{equation*}
    \mathbf{P}(\hat{f}^e_j) = \mathbf{n}^{(F,e)} w_s(\hat{f}^{e,R}_j - \hat{f}^{e,L}_j)/2,
\end{equation*}
where 
$w_s$ is the wave speed across the interface, which depends on the specific equation to be solved. This gives the following equation for $\mathbf{f}^*$:
\begin{equation}
    \mathbf{f}^*_j = \frac{1}{2}\left(\mathbf{\hat{f}}^{e,R}_j + \mathbf{\hat{f}}^{e,L}_j - \mathbf{n}^{F,e}w_s(\hat{f}^{e,R}_j-\hat{f}^{e,L}_j)\right), \quad j=1,\dots,(N+1)^3,
\end{equation}
where $\mathbf{\hat{f}}^e_j = (\hat{f}^e_j,\hat{f}^e_j,\hat{f}^e_j)$.
We note that in the DG formulation for $G(f)=\boldsymbol{\nabla}\cdot f + \boldsymbol{\nabla}^2f$ the boundary term in (\ref{eq:second_derivative}) does not vanish and needs to be evaluated. 
Such term is treated in a similar fashion as the boundary term in \eqref{eq:53}. For the details, we refer the interested reader to
 \cite{giraldoBOOK,hesthavenWarburton2008}.

\subsection{Non-column based rain sedimentation}\label{sec:no_column_rain}

The main novelty of this work lies in the computation of the sedimentation term for the rain equation (i.e., the last term on the right-hand side in Eq.~\eqref{eq:qr_1}) which differs from the methods in, e.g., \cite{kessler1969,klempWilhelmson1978,soongOgura1973,oguratakahashi,houze1993book}. The typical column-based approach to handle the sedimentation term is by computing the spatial derivative along each individual column starting from the top of the domain and descending. See, e.g., \cite{gabersekGiraldoDoyle2012,marrasEtAl2013b} for a spectral element implementation of this approach.

Although widely used, the traditional column-based implementation has a main drawback:
it requires the availability of column-aware data structures that may not serve other purposes in the numerical method, thereby forcing the use of structured grids. 
Unstructured grids are highly advantageous around topography. By forgoing the use of columns, our approach to compute sedimentation could help yield more accurate predictions for storm behavior in mountainous regions. 

Computing the sedimentation term is done separately from the other microphysics calculations, and is done after solving the compressible Euler and moisture advection equations. This term is included by solving the following equation:
\begin{equation}\label{sed}
    \frac{\partial q_r}{\partial t} = \frac{1}{\rho} \frac{\partial}{\partial z}(\rho q_r w_r)
\end{equation}
in non-conservative form and
\begin{equation}\label{sed_c}
    \frac{\partial (\rho q_r)}{\partial t} = \boldsymbol{\nabla} \cdot (\rho q_r w_r \mathbf{\hat{k}})
\end{equation}
in conservation form. Given that $\mathbf{\hat{k}}=(0,0,-1)^T$ for the domains we consider, \eqref{sed_c} can be written as follows:
\begin{equation*}
   \frac{\partial (\rho q_r)}{\partial t} = \frac{\partial}{\partial z}(\rho q_r w_r).
\end{equation*}
This makes it so that for either the conservative or non-conservative form, solving the sedimentation equation essentially amounts to calculating the term $\frac{\partial}{\partial z}(\rho q_r w_r)$.

We can rewrite the sedimentation equation in the form of \eqref{eq:abstraction_law} by taking 
$G(f) = -c\frac{\partial F_{\rm sed}}{\partial z}$, where $F_{\rm sed}=(\rho q_r w_r)$, $c=1$ and
$f=\rho q_r$ in conservation form, while $c=\frac{1}{\rho}$ and $f=q_r$ in non-conservative form. By multiplying by the expansion functions and integrating, we get:
\begin{equation*}
    \int_{\Omega_e}\psi_i(\mathbf{x})\frac{\partial f^h(\mathbf{x},t)}{\partial t} d\mathbf{x} = \int_{\Omega_e}\psi_i(\mathbf{x})\sum_{j=1}^{(N+1)^3}\frac{\partial \psi_j(\mathbf{x})}{\partial z}c\hat{F}^e_{j,sed}(t)(\mathbf{x})d\mathbf{x}, \quad i=1,\dots,(N+1)^3,
\end{equation*}
where $\hat{F}_{j,sed}$ are the expansion coefficients of ${F}_{j,sed}$. Moving to the reference element and identifying the mass matrix yields
\begin{equation}\label{eq:prev}
    \mathbf{M}_{ij}^e \frac{\partial \hat{f}^e_j(t)}{\partial t} = \int_{\Omega_{ref}} \psi(\boldsymbol{\xi})\left[\boldsymbol{\nabla}_{\xi} \psi_j(\boldsymbol{\xi})\cdot \left(\frac{\partial \xi}{\partial z}, \frac{\partial \eta}{\partial z}, \frac{\partial \zeta}{\partial z}\right)(\boldsymbol{\xi})\right] c\hat{F}^e_{j,sed}(t) |\mathbf{J}(\boldsymbol{\xi})|d\boldsymbol{\xi},
\end{equation}
where $i,j=1,\dots,(N+1)^3$. Let us call $\mathbf{D}^e_{\rm sed}$ the element-wise differentiation matrix for \eqref{sed} and write Eq.~\eqref{eq:prev} in matrix form:
\begin{equation}
    \mathbf{M}^e_{ij} \frac{\partial \hat{f}^e_j(t)}{\partial t} = \mathbf{D}^e_{ij,{\rm sed}}c\hat{F}^e_{j,sed}(t),\quad i,j=1,\dots,(N+1)^3.
\end{equation}
We can write $\mathbf{D}^e_{\rm sed}$ discretely as follows:
\begin{align}
    \mathbf{D}^e_{ij,{\rm sed}} = \sum_{k=1}^{N+1}\sum_{m=1}^{N+1}\sum_{n=1}^{N+1}\omega(\xi_k,\eta_m,\zeta_n)\psi_i(\xi_k,\eta_m,\zeta_n)\boldsymbol{\nabla}_{\xi} \psi_j(\xi_k,\eta_m,\zeta_n)
    \cdot\left(\frac{\partial \xi}{\partial z},\frac{\partial \eta}{\partial z},\frac{\partial \zeta}{\partial z}\right)(\xi_k,\eta_m,\zeta_n)|\mathbf{J}(\xi_k,\eta_m,\zeta_n)|,
\end{align}
where $i,j=1,\dots,(N+1)^3$. From this point, if CG is used the global equation can be solved using DSS as follows:
\begin{equation}
    \frac{\partial \mathbf{f}^h}{\partial t} - \mathbf{M}^{-1}\mathbf{D}_{\rm sed}(\mathbf{c} \odot \mathbf{F}^h_{\rm sed}) = 0,
\end{equation}
where $\mathbf{D}_{\rm sed} = \sum_{e=1}^{N_e}\mathbf{D}^e_{\rm sed}$, $\mathbf{c}$ is the vector containing the nodal values of c, $\mathbf{F}_{\rm sed}$ is the vector containing the nodal values of $F_{\rm sed}$, and $\odot$ denotes a component-wise multiplication.
The local DG problem is given by:
\begin{equation}
    \frac{\partial \mathbf{f}^h}{\partial t} - \mathbf{M}^{-1(e)}(\mathbf{\hat{D}}^e_{\rm sed}(\mathbf{c} \odot \mathbf{F}^h_{\rm sed}) - \mathbf{F^e}(\mathbf{c}^* \odot \mathbf{F}^*_{\rm sed})),
\end{equation}
where $\mathbf{\hat{D}}^e_{\rm sed}$ is the weak form of $\mathbf{D}^e_{\rm sed}$, $\mathbf{F}^e$ is the flux matrix at each element, $\mathbf{F}^*_{\rm sed}$ is the interface value of $\mathbf{F}_{\rm sed}$, and $\mathbf{c}^*$ is the interface value of $\mathbf{c}$.

In what follows, we present the procedure we use to solve the fully compressible Euler equations with moisture, including rain.
Algorithm 1 summarizes the entire procedure. The algorithm makes use of the following quantities: $N_{LGL} = N+1$ is the number LGL points in each element, $\Delta t$ is the time step, $f^{h,n}$ is the approximation of $f^h$ at the time $t^n=n\Delta t$, $N_{points}$ the total number of points the domain has been discretized into including repeating nodes at element edges and faces, $t_n$ the current discrete time, and $t_{n+1} = t_n +\Delta t$ . We also define the sedimentation Courant number, which we use to determine the time sub-step for the sedimentation problem:
\begin{equation}\label{eq:Cr-sed}
Cr = w_r \frac{\Delta t}{\Delta z}. 
\end{equation}
This number is used to determine the appropriate sedimentation time step as follows:
\begin{equation}\label{eq:substep}
    \Delta t_{\rm sed} = \frac{\Delta t}{\max(1, 0.5+Cr_{\rm max}/Cr_{\rm limit})},
\end{equation}
where:
\begin{equation}\label{Crmax}
  Cr_{\rm max} = \max(\left[Cr_{i}\right]_{i=1}^{N_{points}}), 
\end{equation} is the maximum sedimentation Courant number among all points in the domain and $Cr_{\rm limit}$ is the maximum allowable Courant number for the sedimentation problem.
The rest of the notation is defined in Sec.~\ref{sec:num_meth}.

\begin{algorithm}
	\caption{Simulation of moist-air and rain sedimentation with unstructured grids.}
	\begin{algorithmic}[1]
		\For {$time =0,\Delta t,\ldots, t_f$}
			\For {$e=1,2,\ldots,N_e$}
			    \For{$node=1,2,\ldots,N_{LGL}$}
			    \State Calculate contributions to element-wise derivatives from each LGL point along 
			    \State the reference element. 
			    \EndFor
			    \State Compute these local derivatives in physical space. 
			\EndFor
	    \State Perform DSS for CG or calculate numerical fluxes for DG.
	    \State Solve the discrete version of the Euler equations: \eqref{eq:mass}, \eqref{eq:mom_c} and \eqref{eq:ent_c} if using conservation \State form, and \eqref{eq:mass}, \eqref{eq:mom} and \eqref{eq:ent} if using non-conservative form.
	    \State Solve the advection equations for $q_v$, $q_c$ and $q_r$ by the flow velocity $u$: \eqref{eq:moist1_0}, \eqref{eq:moist2_0} and \State \eqref{eq:qr_0} if using conservation form, and \eqref{eq:moist1_1}, \eqref{eq:moist2_1} and \eqref{eq:qr_1} if using non-conservative form.
	    \For {$i=1,2,\dots,N_{points}$}
	        \State Determine $w_r$  using Eq.~(\ref{terminal_velocity}) 
	        \State Determine $Cr_{\rm max}$ using \eqref{Crmax}
	        \State Determine $\Delta t_{\rm sed}$ using \eqref{eq:substep}
	    \EndFor
	        \For{ts} = $t_n$,$t_n+\Delta t_{\rm sed}$,\dots,$t_{n+1}$
	            \For {$e=1,2,\ldots,Ne$}
	            \State \textbf{if} space method == CG
	                \State ~~~~~ Compute $\mathbf{D}^e_{\rm sed}$
	            \State \textbf{else if} space method == DG
	                \State ~~~~~Compute $\mathbf{\hat{D}}^e_{\rm sed}$
	            \State \textbf{end if}
		        \EndFor
		        \State \textbf{if} space method == CG
		            \State ~~~~~Perform DSS.
		        \State \textbf{else if} space method == DG
		          \State ~~~~~Apply inter-element fluxes for the sedimentation equation using $w_r$ as the wave \State ~~~~~speed.
		        \State \textbf{end if}
	            \State Solve \eqref{sed}
	        \EndFor
	        \For {$e=1,2,\ldots,Ne$} 
	        \State Update moisture variables and potential temperature to account for phase changes 
	        \State following equations \eqref{tempeq}-\eqref{raineq}
	        \EndFor
	   \EndFor
	\end{algorithmic} 
\end{algorithm}

Next, we report on the results obtained with this algorithm and fully unstructured grids. 
\section{Results}\label{sec:num_res}

We assess the method presented in Sec.~\ref{sec:no_column_rain} with an idealized squall line test from \cite{gabersekGiraldoDoyle2012} and a fully 3D supercell problem from \cite{skamarockEtAl2012}. All the simulations are run with the Nonhydrostatic Unified Model of the Atmosphere (NUMA) \cite{kellyGiraldo2012}, which is designed to solve the dry Euler equations, with the addition of artificial viscosity as described in Sec.~\ref{sec:num_meth}, on unstructured grids of hexahedra with arbitrary orientation. NUMA enables the use of both CG and DG spectral elements and has been shown to scale exceptionally well on CPUs and GPUs  in \cite{abdiEtAl2017a,Muller2018}.

\subsection{2.5D Squall line}\label{sec:squall_line}

The first benchmark we consider is an idealized test presented in \cite{gabersekGiraldoDoyle2012}.  
While the computational domain in \cite{gabersekGiraldoDoyle2012} is two-dimensional, we run the same test in a $2.5$~D domain 
$\Omega = [150 \times 12 \times 24$] ${\rm km^3}$. The domain
is discretized with a single element in the $y$ direction and a resolution dependent number of elements in the $x$ and $z$ directions.
Periodic boundary conditions are applied to the lateral boundaries, a free-slip type boundary condition is applied at the domain bottom and the domain top utilizes a Rayleigh sponge for gravity wave damping.
In this domain, a squall line forms in a weakly stable atmosphere with Brunt-V\"{a}is\"{a}l\"{a} frequency $N = 0.01~ {\rm s^{-1}}$ below the tropopause and a more stable atmosphere with $N = 0.02~ {\rm s^{-1}}$ above 12 km. The cloud begins to form around $t\approx500$~s, while rain starts to form and fall at approximately $t\approx900$~s. The initial condition consists of a saturated boundary layer typical of mid-latitude storms that has been used in several numerical studies (see, e.g, \cite{rotunnoKlempWeisman1988, weismanEtAl1988}). A low altitude wind shear in the $x$ direction is imposed to break the cloud symmetry and allow for a continuous storm evolution. The initial background sounding is tabulated in the Appendix. 

The storm is triggered by a thermal perturbation of the background state \cite{rotunnoKlempWeisman1988} centered at $(x_c, z_c)=(75000, 2000)$ m and defined by:
\begin{equation}
    \Delta\theta = 
    \begin{cases}
    \theta_c\cos\left(\frac{\pi r}{2} \right) & \textrm{if $r\leq r_c$}, \\
    0 & \textrm{if $r\geq r_c$},
    \end{cases}
\label{warmEqn}
\end{equation}
where 
\[
r = \sqrt[]{\frac{(x-x_{c})^{2}}{r_x^2} + \frac{(z-z_{c})^{2}}{r_z^2}}, \quad \theta_c = 3~\text{K}, \quad r_c= 1, \quad r_x = 10000~\text{m}, \quad r_z = 1500~\text{m}.
\]

We generated seven grids using GMSH \cite{GMSH}. 
Table \ref{tab:table1} lists the total number of hexahedral elements and the effective resolution
$\Delta x$ for each mesh. We choose to report the effective resolution because the LGL points for an element are not equidistant \cite{giraldoBOOK,hesthavenWarburton2008,KoprivaBook}. NUMA relies on P4est \cite{BursteddeWilcoxGhattas11} to read unstructured meshes and perform the graph partitioning for the parallel application.

\begin{figure}[h!]
    \centering
    \includegraphics[width=.8\textwidth]{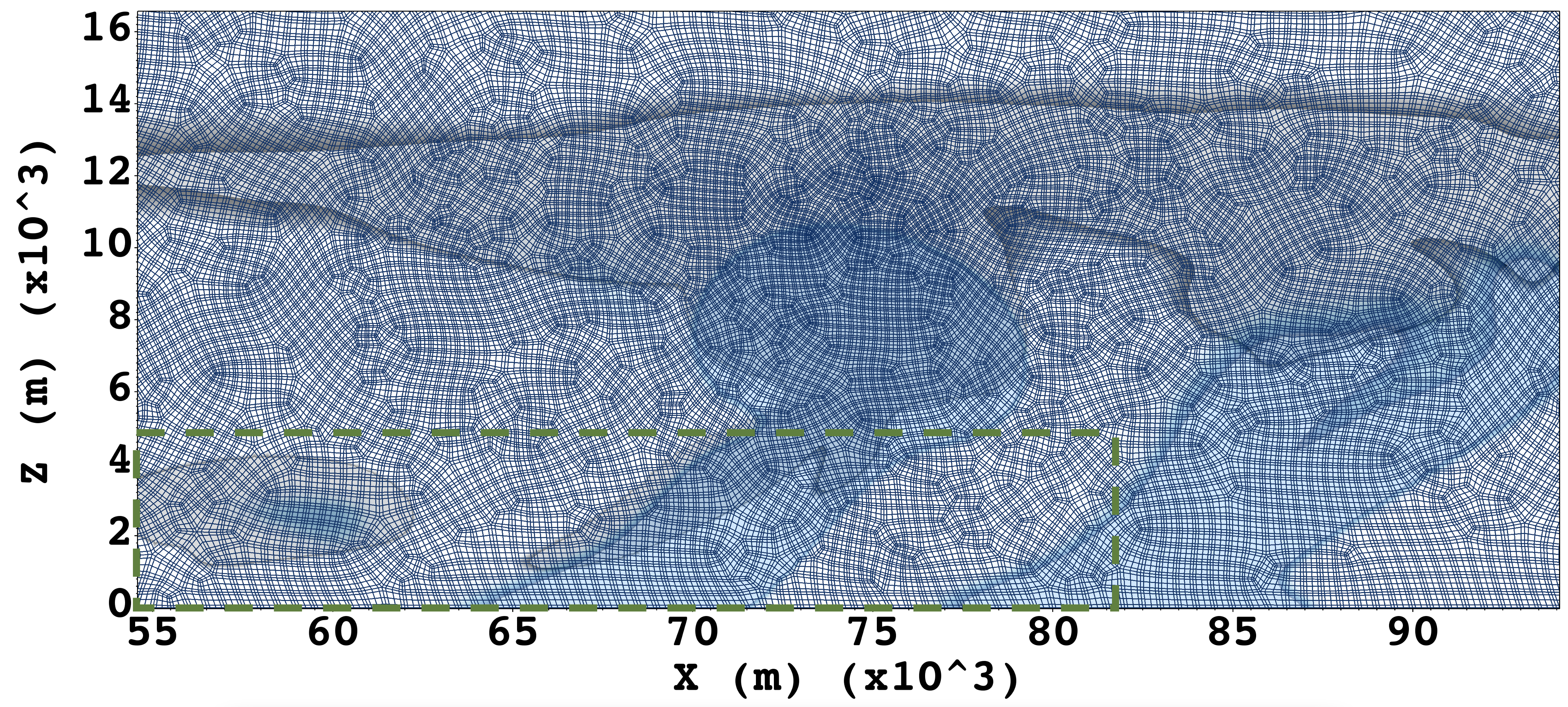}
	\includegraphics[width=.8\textwidth]{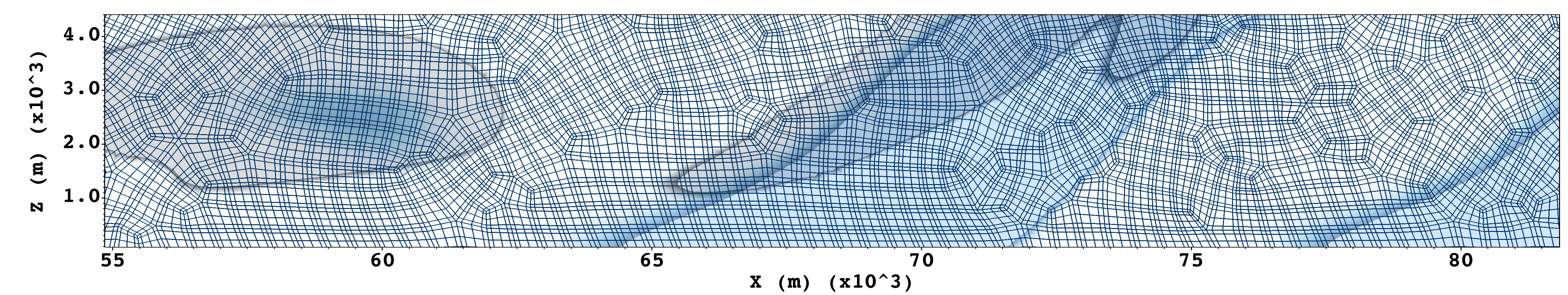}
      \caption{Top: $q_c$ and $q_r$ over unstructured grid $\Delta x = 150$ m. Cloud water is shaded in grey for values of  $q_c > 1\times 10^{-5~}{\rm kg/kg}$ whereas rain is shaded in blue for values of $q_r > 1\times 10^{-4}~{\rm kg/kg}$. Bottom: close-up view corresponding to the dashed rectangle in the top figure.}
      \label{Grids2}
\end{figure}

Fig.~\ref{Grids2} shows an example of clouds and precipitation calculated on a fully unstructured grid of hexahedra for an effective resolution of 150 m in both spatial directions. 

\begin{table}[h!]\label{table1}
    \centering
        \begin{tabular}{|c|c|c|c|c|c|c|c|}
    \hline
       \# elements   & 473 & 1078 & 3181 & 4134 & 6485 & 11447 & 25863 \\
       \hline
       $\Delta x$ & 750~m & 500~m & 290~m & 250~m & 200~m & 150~m & 100~m \\
        \hline
    \end{tabular}
    \caption{Total number of hexahedral elements and effective resolution for all the meshes used for
    the squall line simulations.}
    \label{tab:table1}
\end{table}

For all the simulations, we use an Additive Runge Kutta third order (ARK3) semi-implicit time integrator and elements of polynomial order 4.
We maintain the acoustic Courant number $C \leq 1$ for all the simulations. While the ARK3 time integrator allows for larger acoustic Courant numbers, we limit the time step for the purposes of obtaining a greater deal of accuracy for the higher-resolution simulations. We run this test using both the CG approach with the governing equations in non-conservation form and 
the DG approach with the governing equations in conservation form. 
Consistently with \cite{gabersekGiraldoDoyle2012}, a constant artificial viscosity of $\beta = 200$ (for the units see Remark \ref{rem1}) is used to stabilize the simulations.

\begin{figure}[h!]
    \begin{tabular}{ll}
    \includegraphics[width=0.4\textwidth]{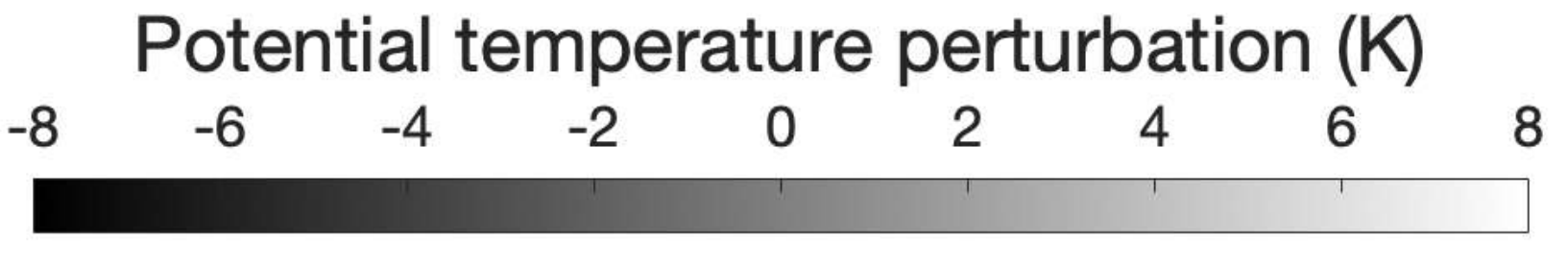} &
    \includegraphics[width=0.4\textwidth]{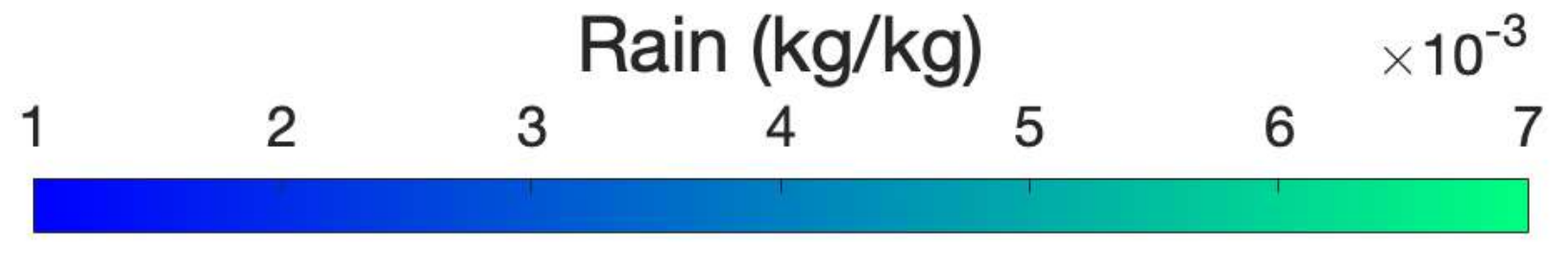}\\
      \includegraphics[width =0.47\textwidth]{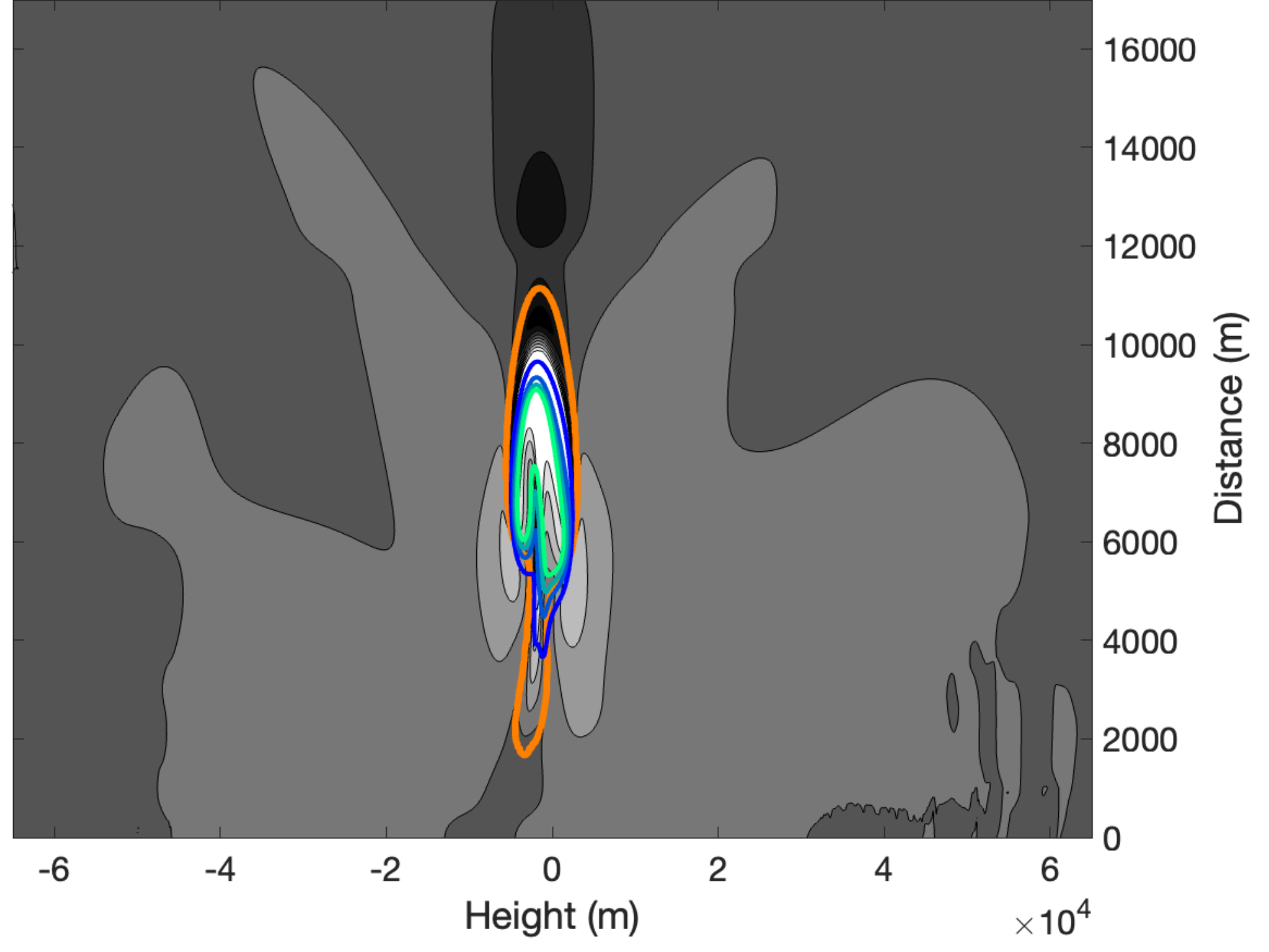}   & \includegraphics[width =0.47\textwidth]{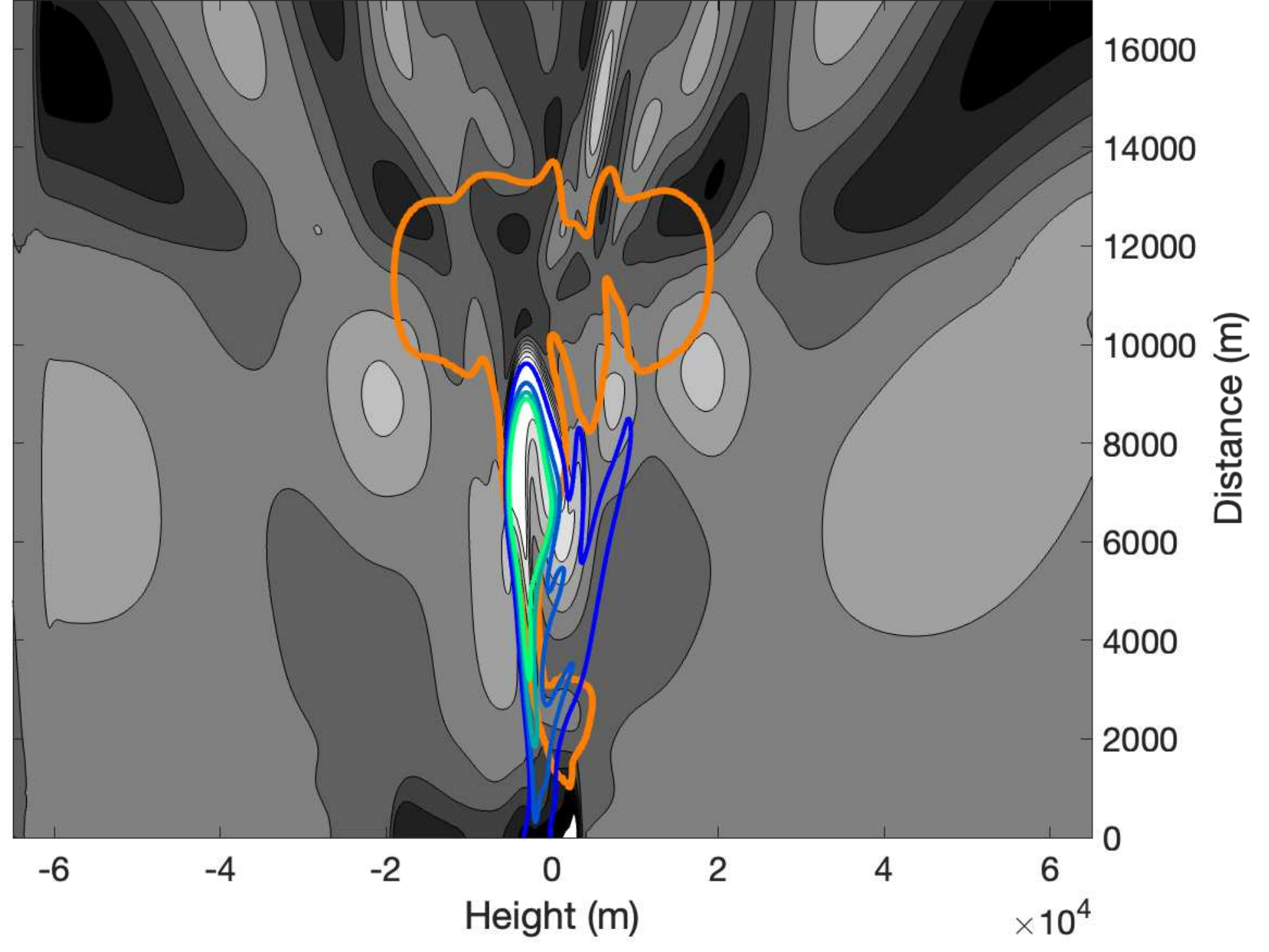} \\
      \includegraphics[width=0.448\textwidth]{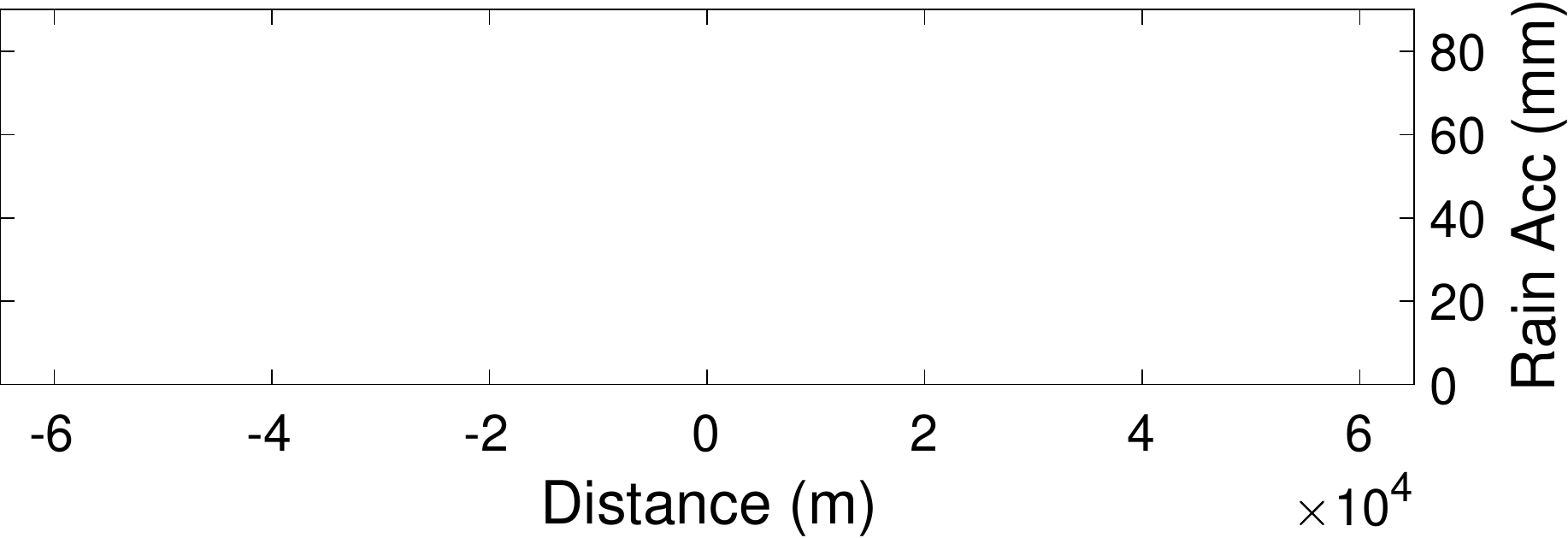} &  \includegraphics[width=0.448\textwidth]{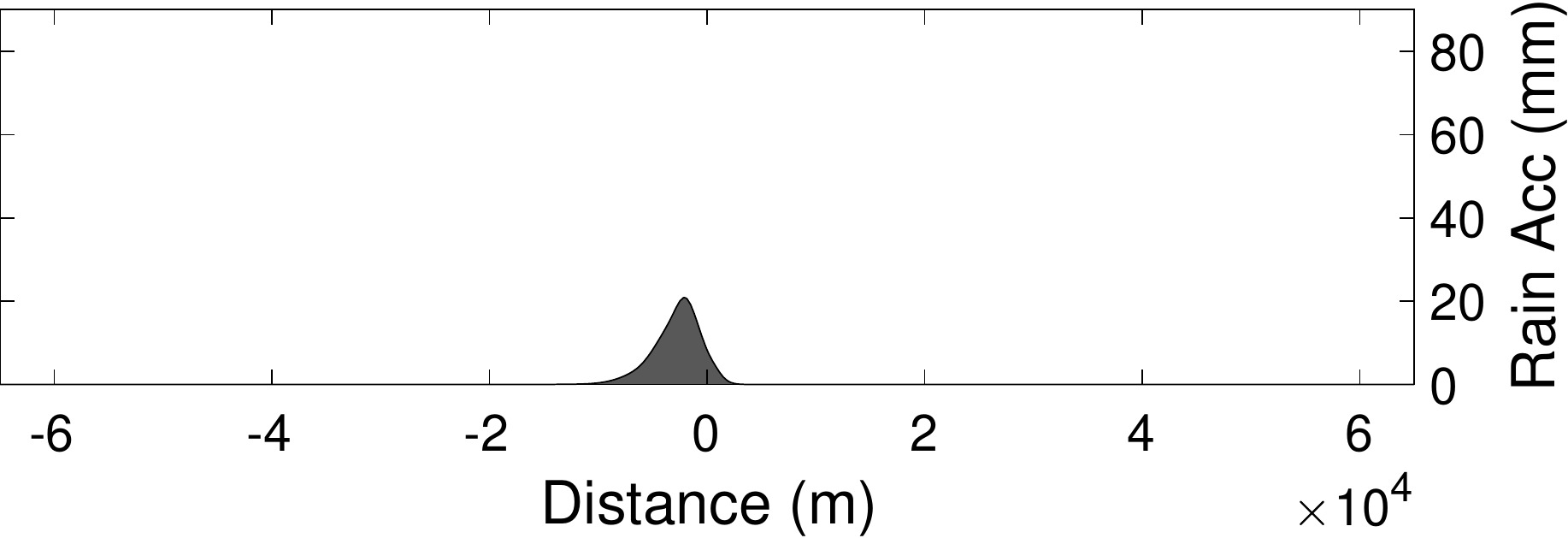} \\
       \includegraphics[width =0.47\textwidth]{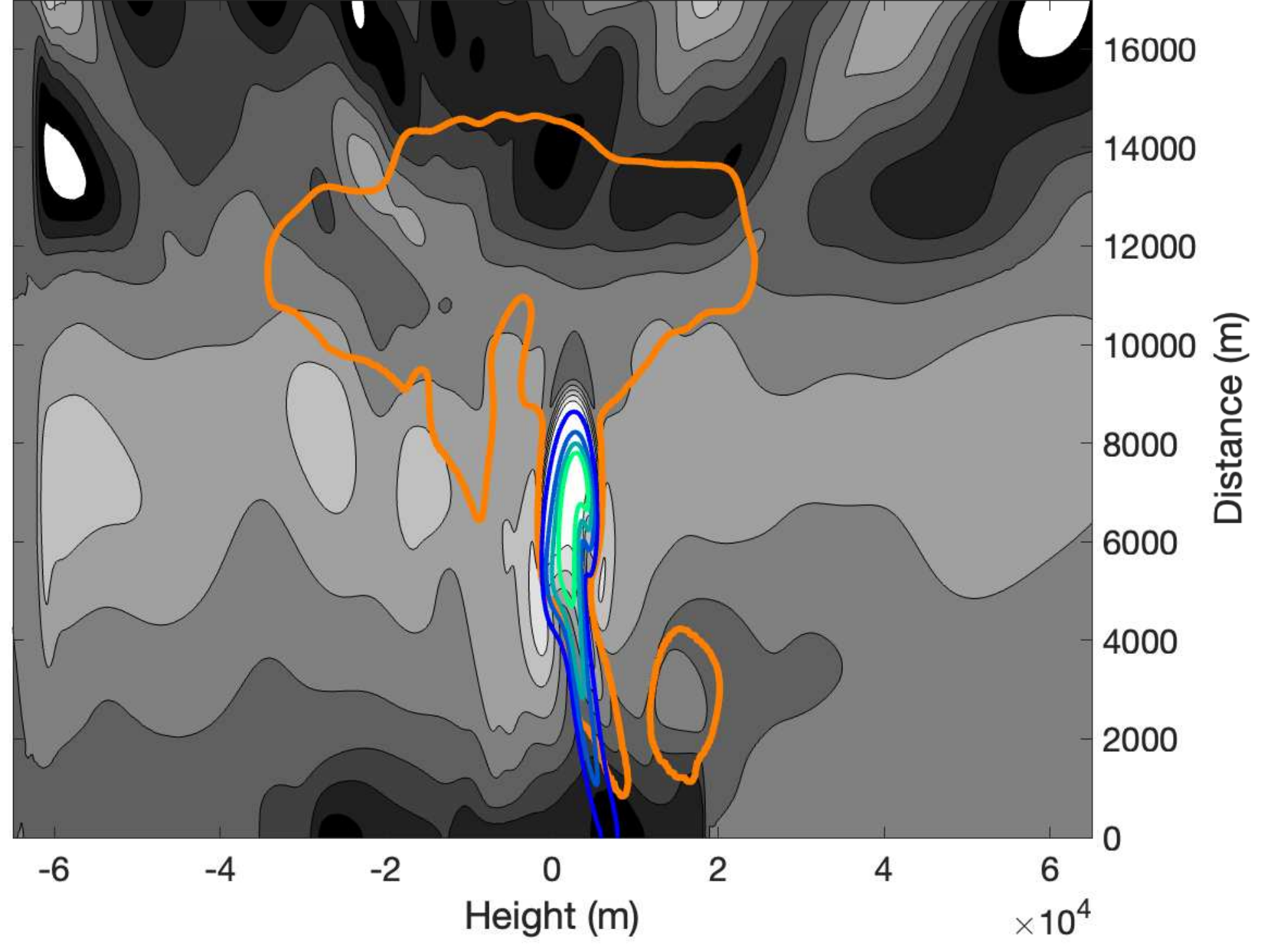}  &  \includegraphics[width =0.47\textwidth]{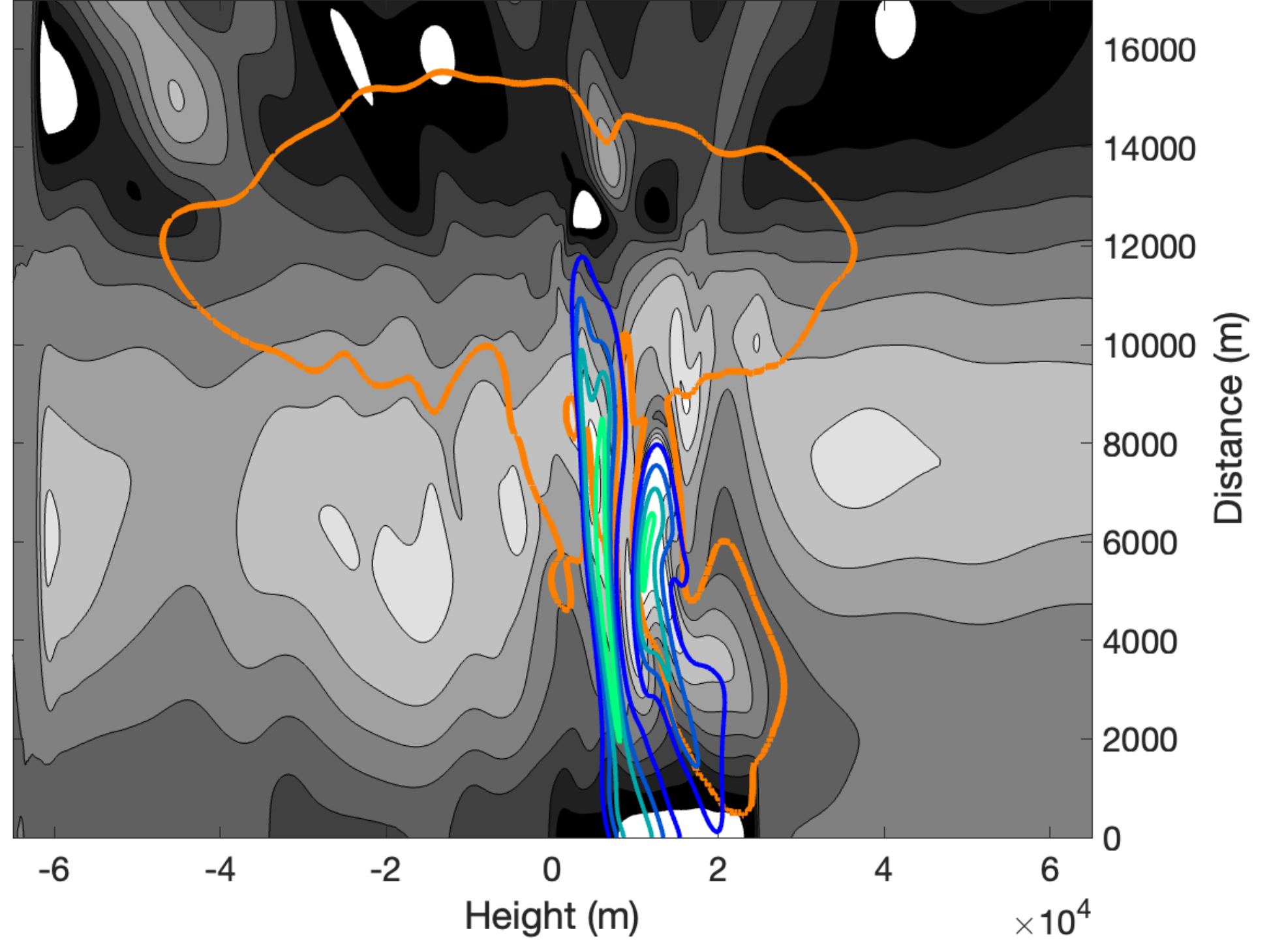}\\
       \includegraphics[width=0.448\textwidth]{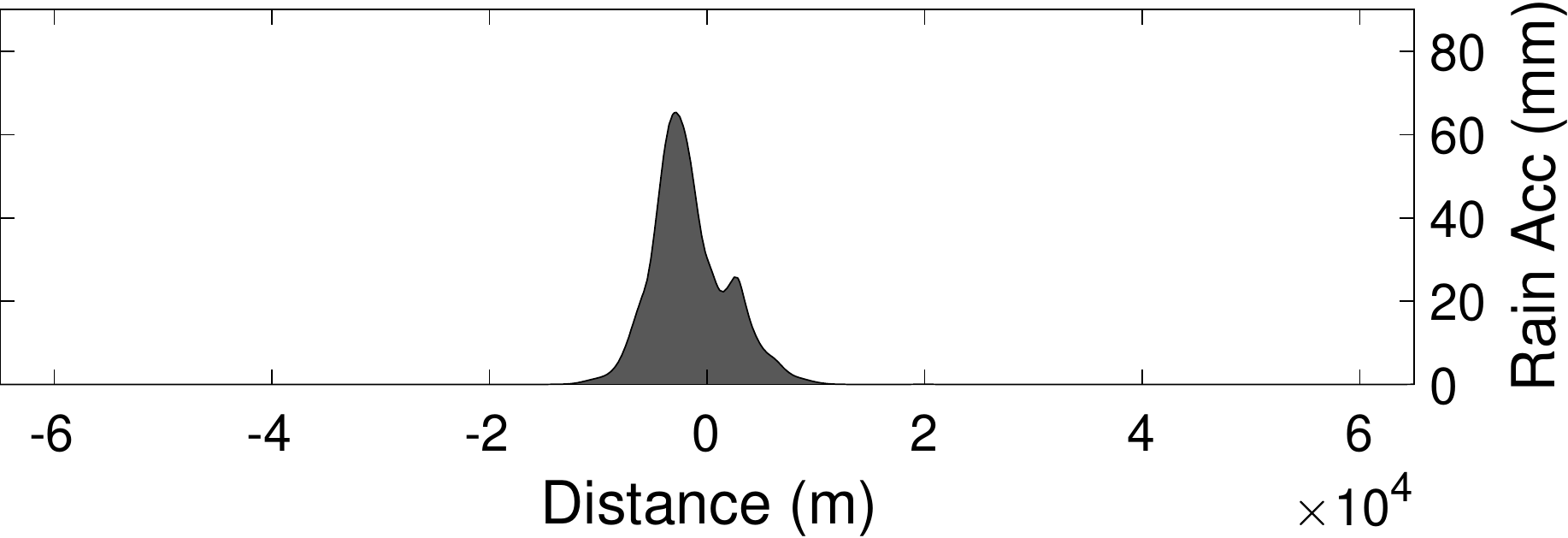} &  \includegraphics[width=0.448\textwidth]{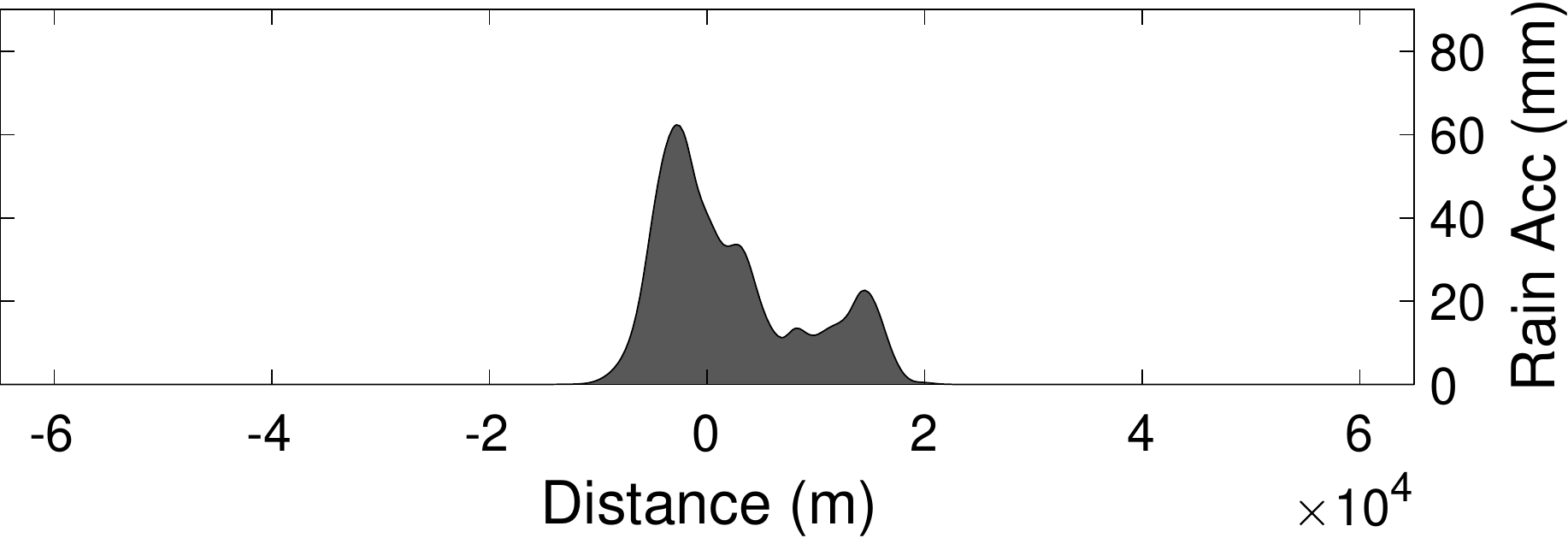}
    \end{tabular}
    \caption{Storm evolution obtained with a CG approximation and mesh with resolution $\Delta x = 100$~m at $t=$1500~s (top-left), 3000~s (top-right), 6000~s (bottom-left) and 9000~s (bottom-right). In the top portion of each panel, the thick orange contour line ($q_c= 10^{-5}~{\rm kgkg^{-1}}$) represents the outline of the cloud. The white and gray contours represent the perturbation potential temperature, and the blue and green contours represent $q_r$. The bottom portion of each panel shows the rain accumulated at the surface for each time as a function of horizontal distance from the point $x=0$~m.}
    \label{storm evolution CG}
\end{figure}

\begin{figure}[h!]
    \begin{tabular}{ll}
    \includegraphics[width=0.4\textwidth]{figures/pot_temp_colorbar.pdf} &
    \includegraphics[width=0.4\textwidth]{figures/cloud_scale_rain.pdf}\\
      \includegraphics[width =0.47\textwidth]{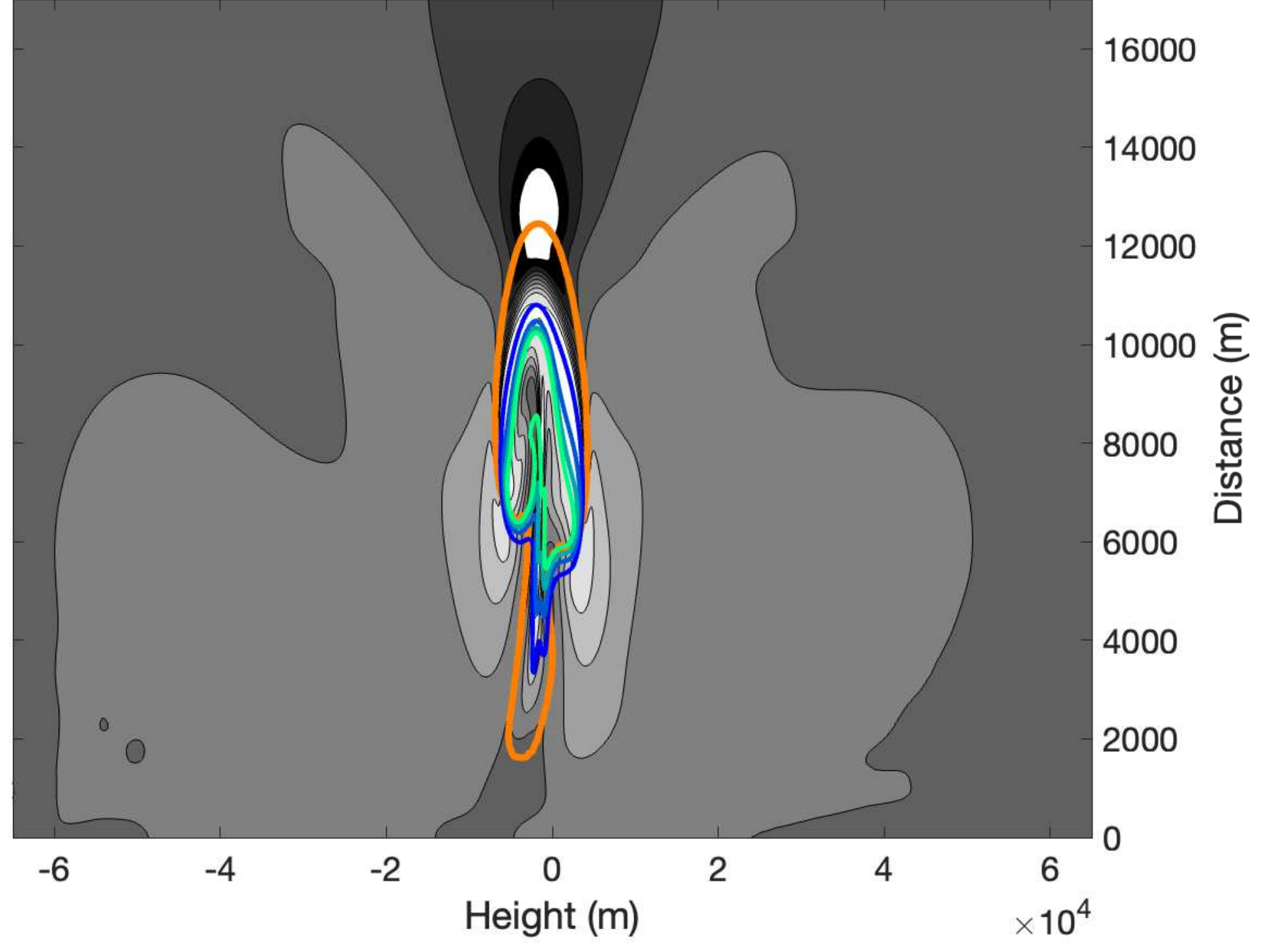}   & \includegraphics[width =0.47\textwidth]{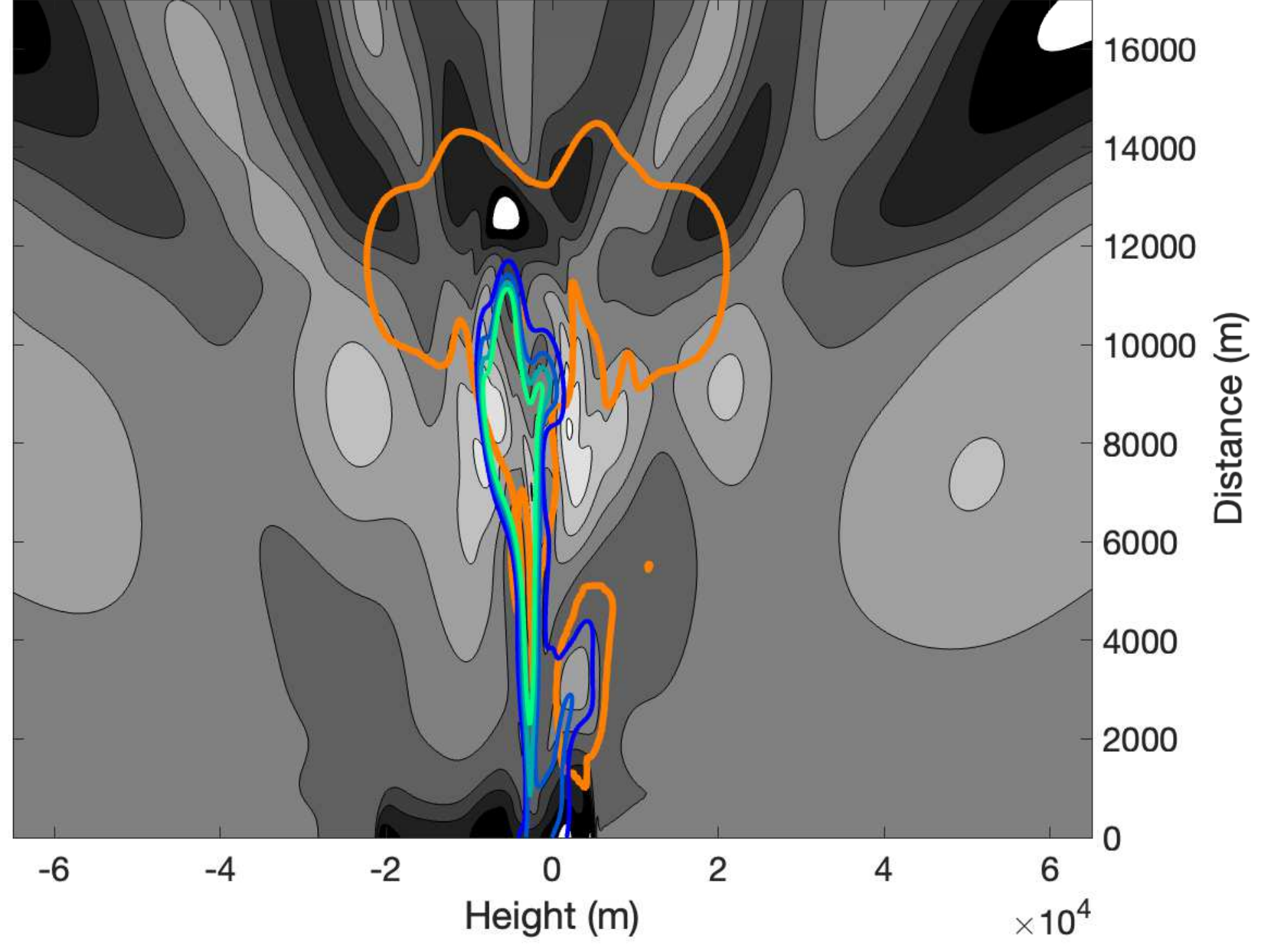} \\
      \includegraphics[width=0.448\textwidth]{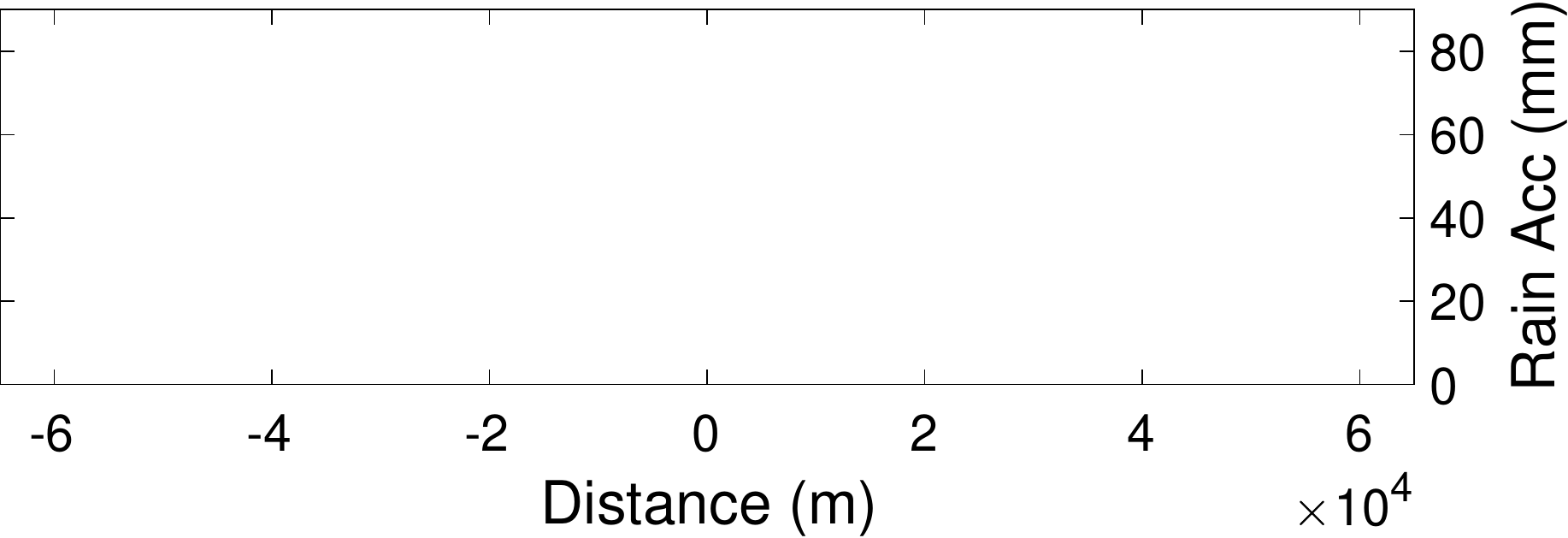} &  \includegraphics[width=0.448\textwidth]{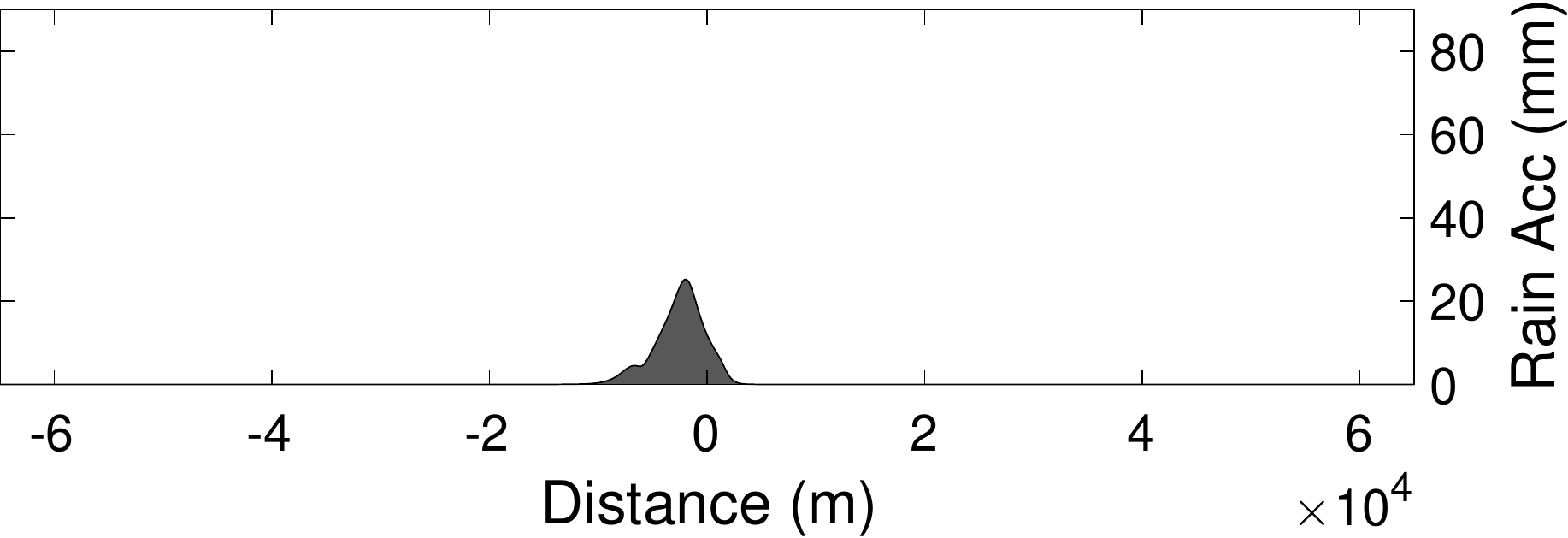} \\
       \includegraphics[width =0.47\textwidth]{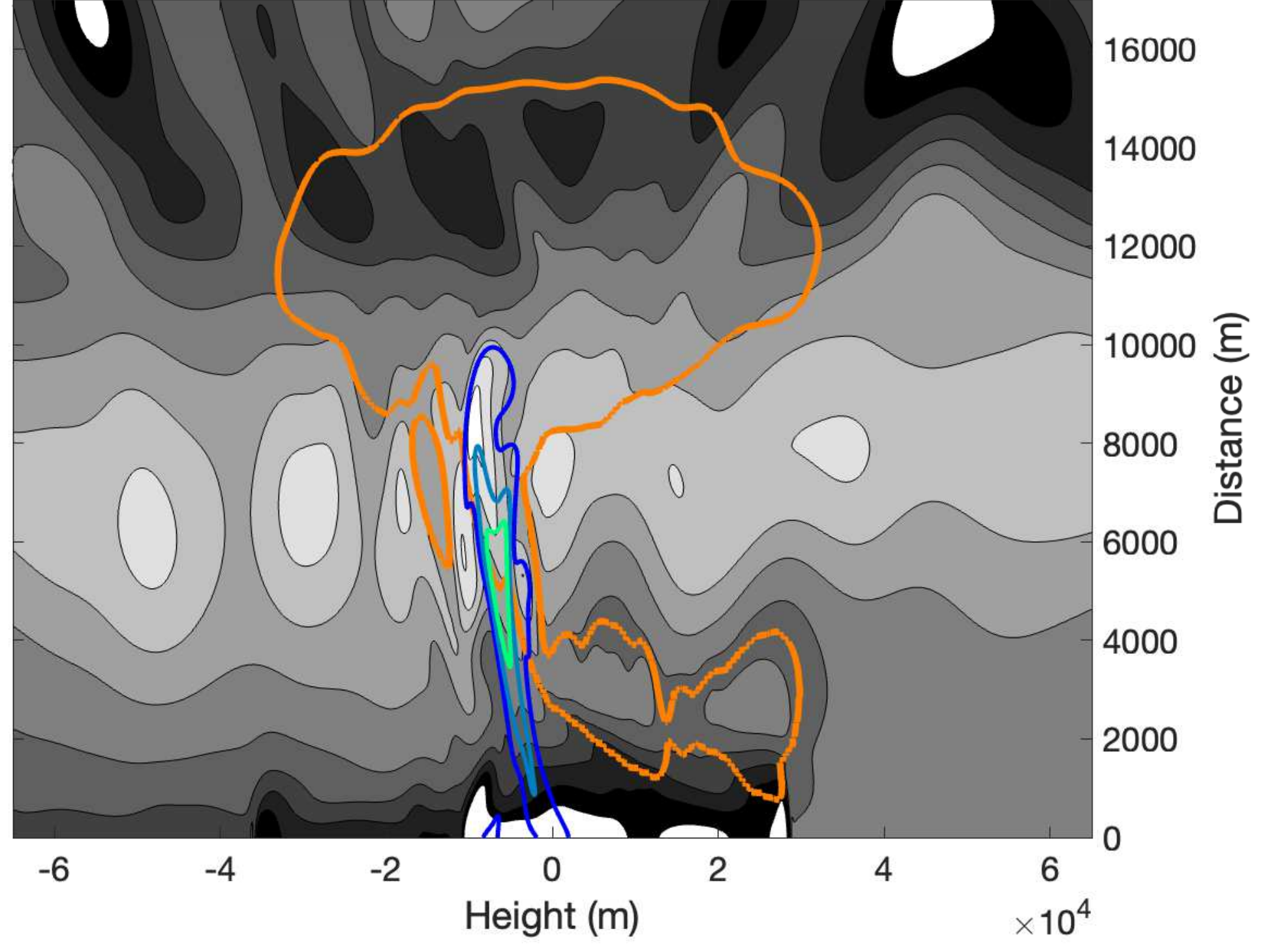}  &  \includegraphics[width =0.47\textwidth]{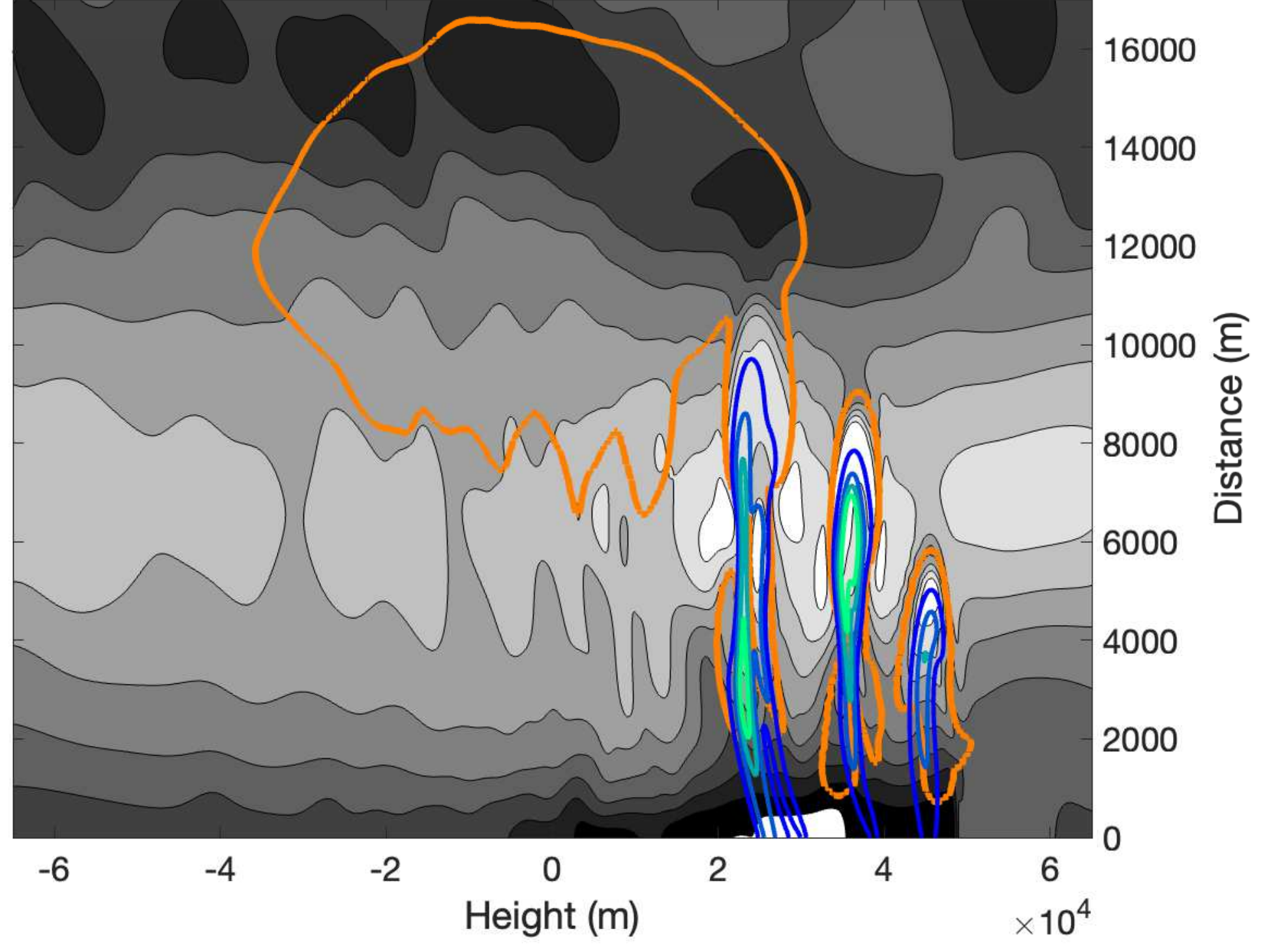}\\
       \includegraphics[width=0.448\textwidth]{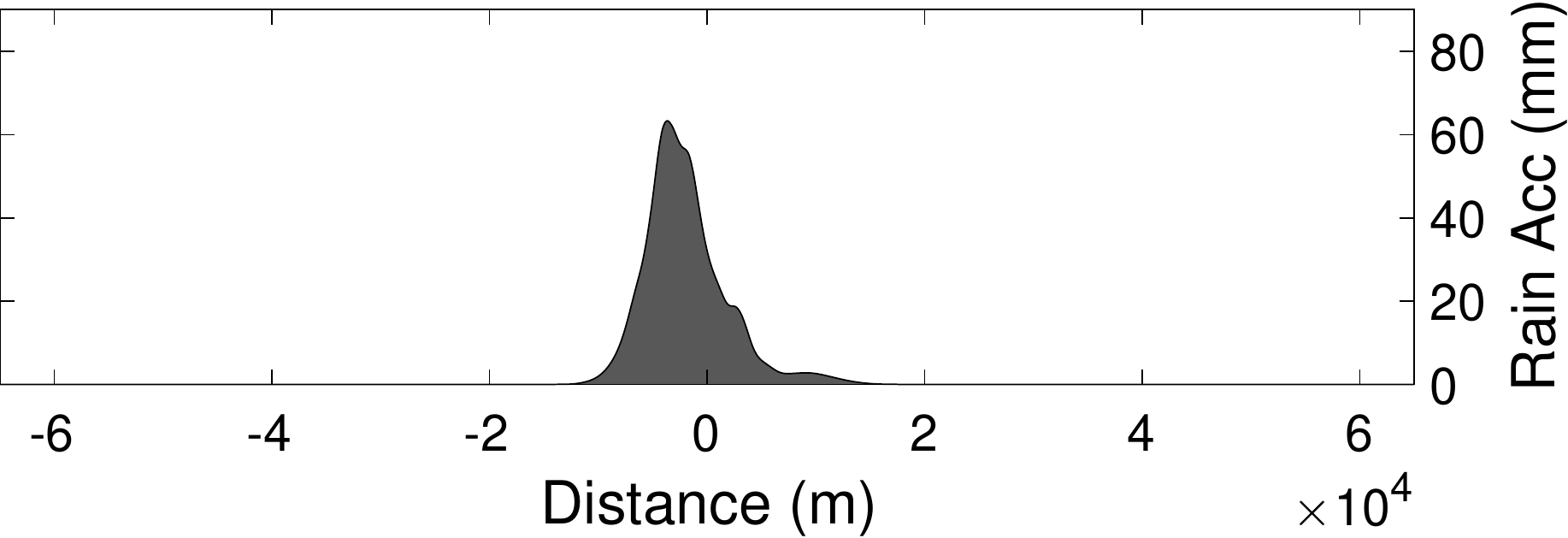} &  \includegraphics[width=0.448\textwidth]{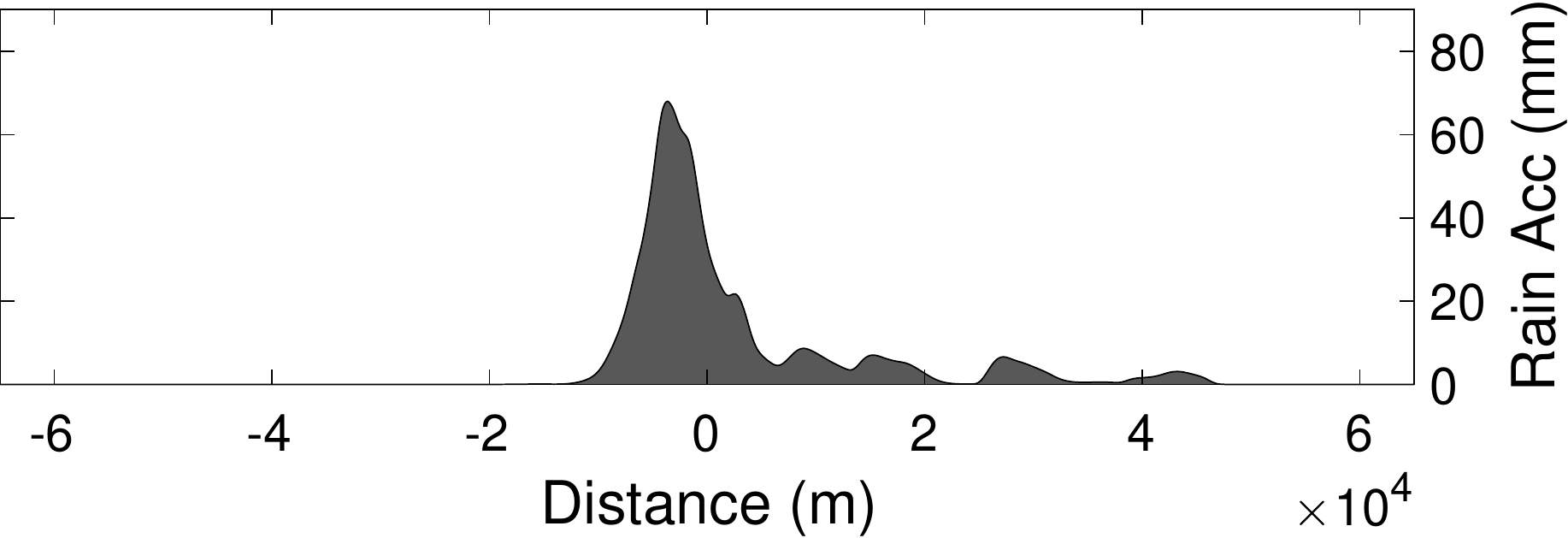}
    \end{tabular}
    \caption{Storm evolution obtained with a DG approximation and mesh with resolution $\Delta x = 100$~m at $t=$1500~s (top-left), 3000~s (top-right), 6000~s (bottom-left) and 9000~s (bottom-right). In the top portion of each panel, the thick orange contour line ($q_c= 10^{-5}~{\rm kgkg^{-1}}$) represents the outline of the cloud. The white and gray contours represent the perturbation potential temperature and the blue and green contours represent $q_r$. The bottom portion of each panel shows the rain accumulated at the surface for each time as a function of horizontal distance from the point $x=0$~m.}
    \label{storm evolution DG}
\end{figure}

Let us examine the results obtained with the finest mesh, i.e.~the one with $\Delta x=100$~m. Figs.~\ref{storm evolution CG} and ~\ref{storm evolution DG} show the stages of the storm evolution given by the CG and DG simulations, respectively. Both simulations yield very similar plots at $t=1500~$s. Additionally, in both cases we observe 
a downwind tilt of the convective tower, which is caused by the horizontal wind-shear, and
the eventual development of the anvil cloud near the tropopause where the atmosphere presents higher stability.
For the sake of brevity, we do not report the plots associated with other meshes, but a similar early storm evolution is observed in all the simulations at all resolutions with both CG and DG approaches.
The differences between the CG and DG simulations remain minimal even up to about $t=6000~$s. This is a rather long
period of time since by then the storm has fully developed.
Starting from $t=6000~$s till the end of the time interval of interest, some differences in the CG and DG simulations
arise, as can been seen by comparing Figs.~\ref{storm evolution CG} and ~\ref{storm evolution DG}.  
At $t=9000~$s, when additional convective towers are observed, the DG simulation generates multiple convective towers, some of which are significantly downwind. This is not as pronounced in the CG simulation. Compare
the bottom right panels in Figs.~\ref{storm evolution CG} and ~\ref{storm evolution DG}.

Figs.~\ref{storm evolution CG} and ~\ref{storm evolution DG} reports also the rain accumulated on the ground. 
At $t = 1500$~s, no rain has accumulated yet in either the DG or CG simulations. This is confirmed by the rain contours plots, where we see that the contour lines have yet to reach the ground. See top left panel in Figs.~\ref{storm evolution CG} and ~\ref{storm evolution DG}. At $t=3000~$s, the accumulated rain is primarily near the center of the domain for both methods. Indeed, from the top right panel in Figs.~\ref{storm evolution CG} and ~\ref{storm evolution DG} we see that rain accumulates at the location of the convective tower, with a slight asymmetry that follows the asymmetry of the convective tower seen at $t = 1500$~s. 
As time progresses, the convective tower tilts. An early stage of this is visible at $t = 3000$~s, but the tilting becomes more pronounced at $t = 6000$~s when the
effect of the wind shear is more noticeable.
The rain accumulation reflects the tilting and location of the convective tower in both the CG and DG simulations, as shown in the bottom left panel of Figs.~\ref{storm evolution CG} and ~\ref{storm evolution DG}.
By $t = 9000$~s, we observe once again some differences in the results given by the two methods. 
For the CG simulation, in the bottom right panel of Fig.~\ref{storm evolution CG} we see a much wider distribution of accumulated rain with a secondary peak below the new location of the convective tower and a third peak appearing below the location of the secondary convective tower. As for the DG simulation, in the bottom right panel of Fig.~\ref{storm evolution DG} we notice that the rain accumulation matches the downwind shifting of the main column and small peaks appear where secondary convective towers are present. 

Regardless of the space discretization method,
we see that once rain appears within the convective tower it is correctly transported downward without the need for a vertically structured grid. This hold true also when
multiple, possibly disconnected, sources of rain are present in the domain. In both sets of simulations, the rain falls to the ground following the location of the convective towers and the effects of the wind-shear. This gives us confidence that our algorithm is able to correctly transport rain despite the lack of a vertically structured grid and regardless of the space discretization method.

\begin{figure}
      \begin{tabular}{ll}
    \includegraphics[width=0.4\textwidth]{figures/pot_temp_colorbar.pdf} &
    \includegraphics[width=0.4\textwidth]{figures/cloud_scale_rain.pdf}
    \\
    \includegraphics[width=0.47\textwidth]{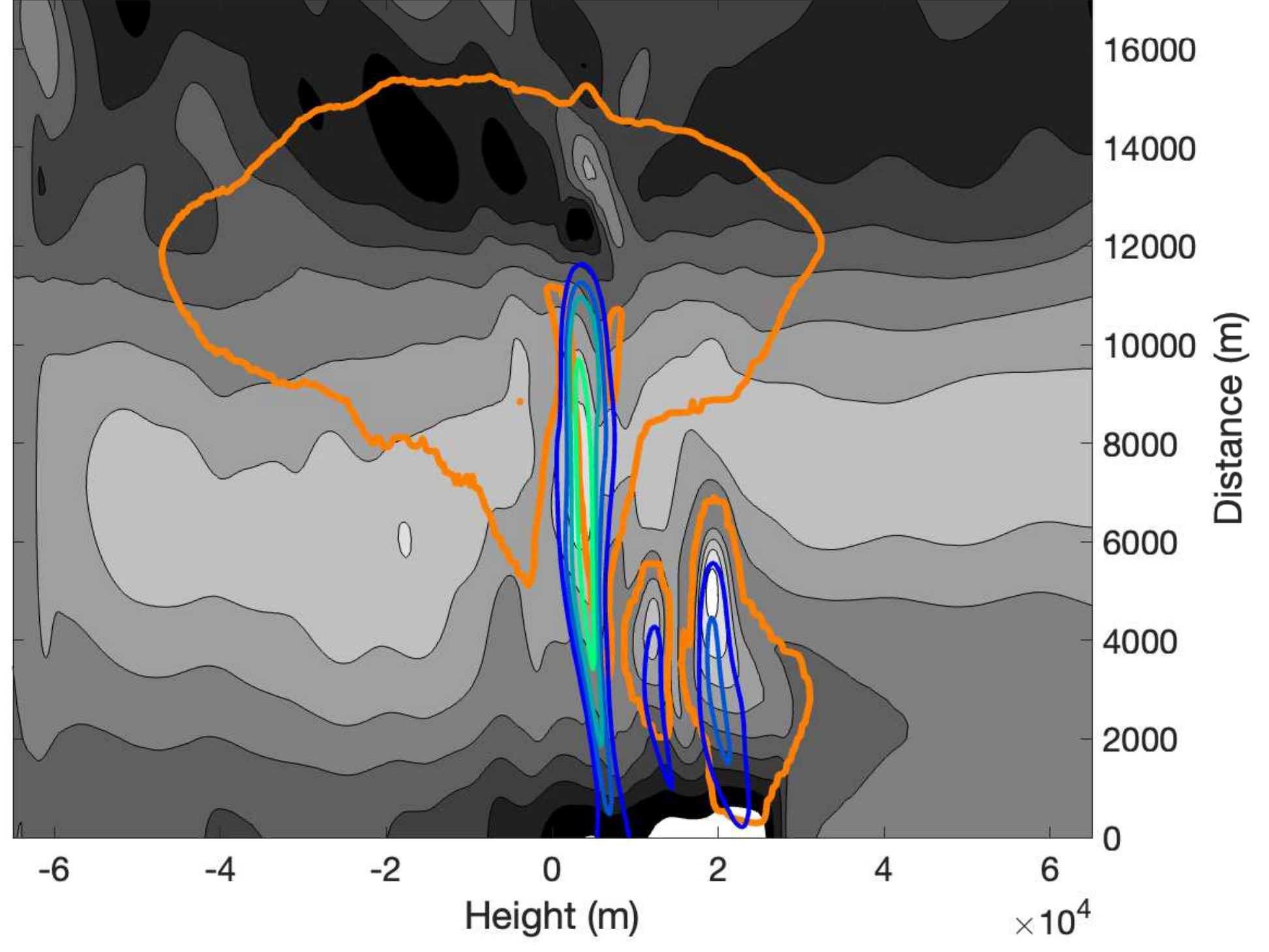} & \includegraphics[width=0.47\textwidth]{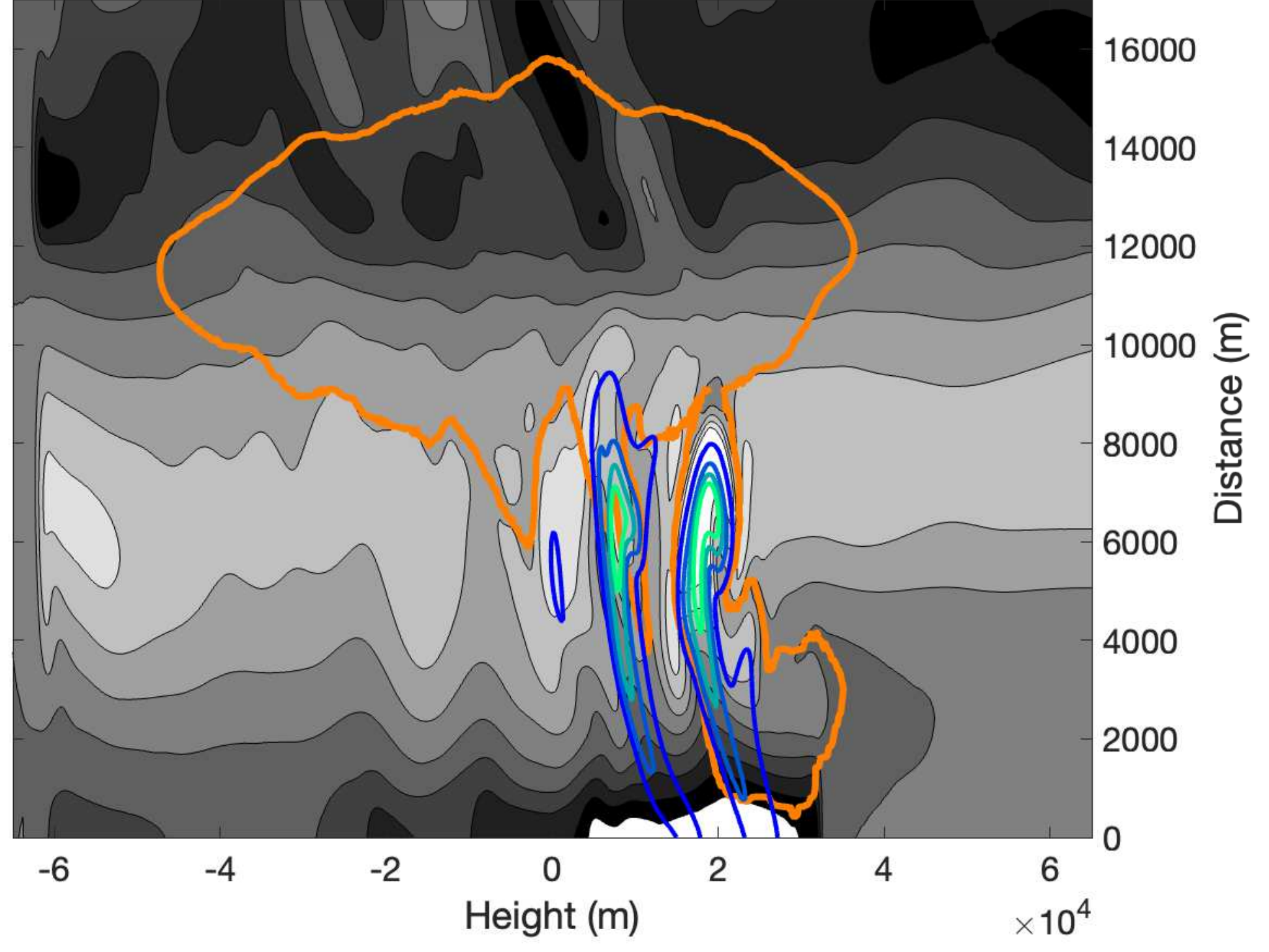} \\
           
            \includegraphics[width=0.448\textwidth]{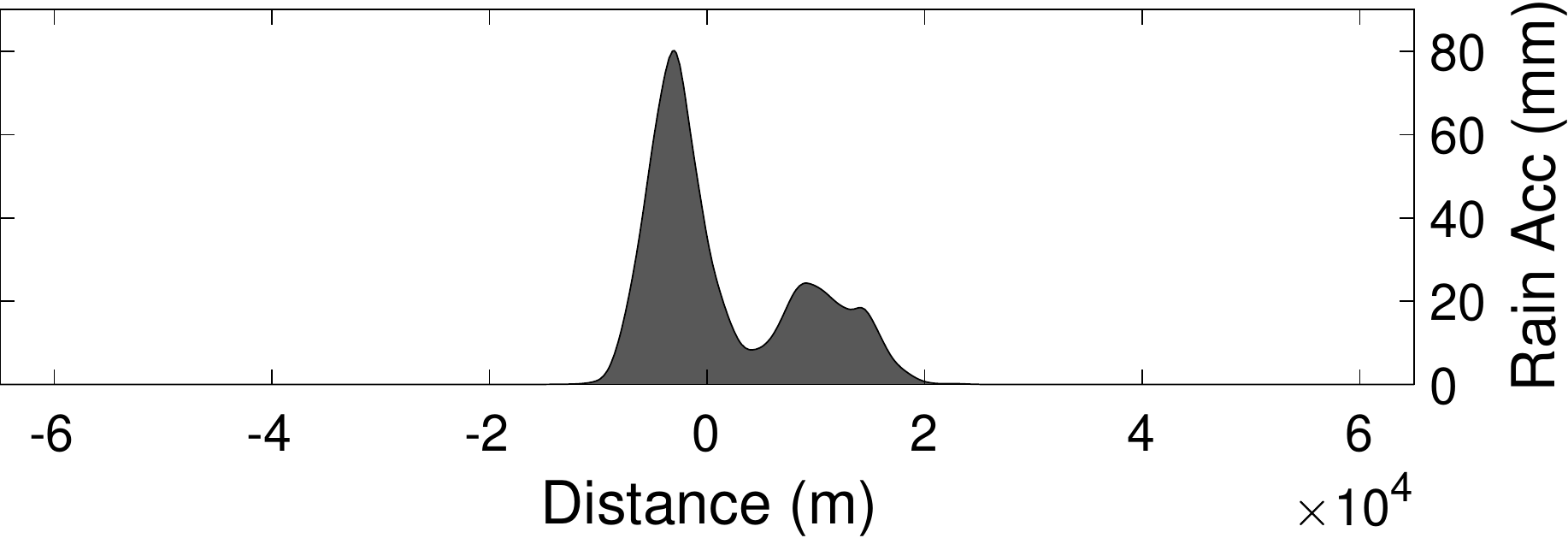} &  \includegraphics[width=0.448\textwidth]{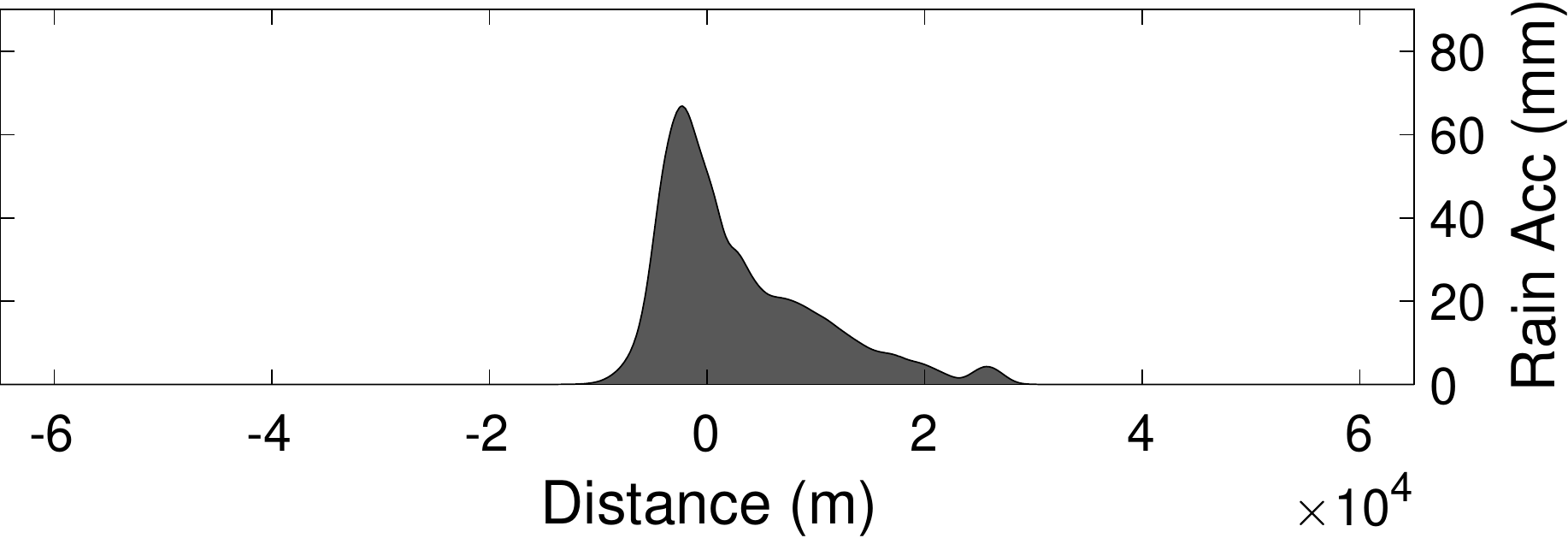}\\
            
            \includegraphics[width=0.47\textwidth]{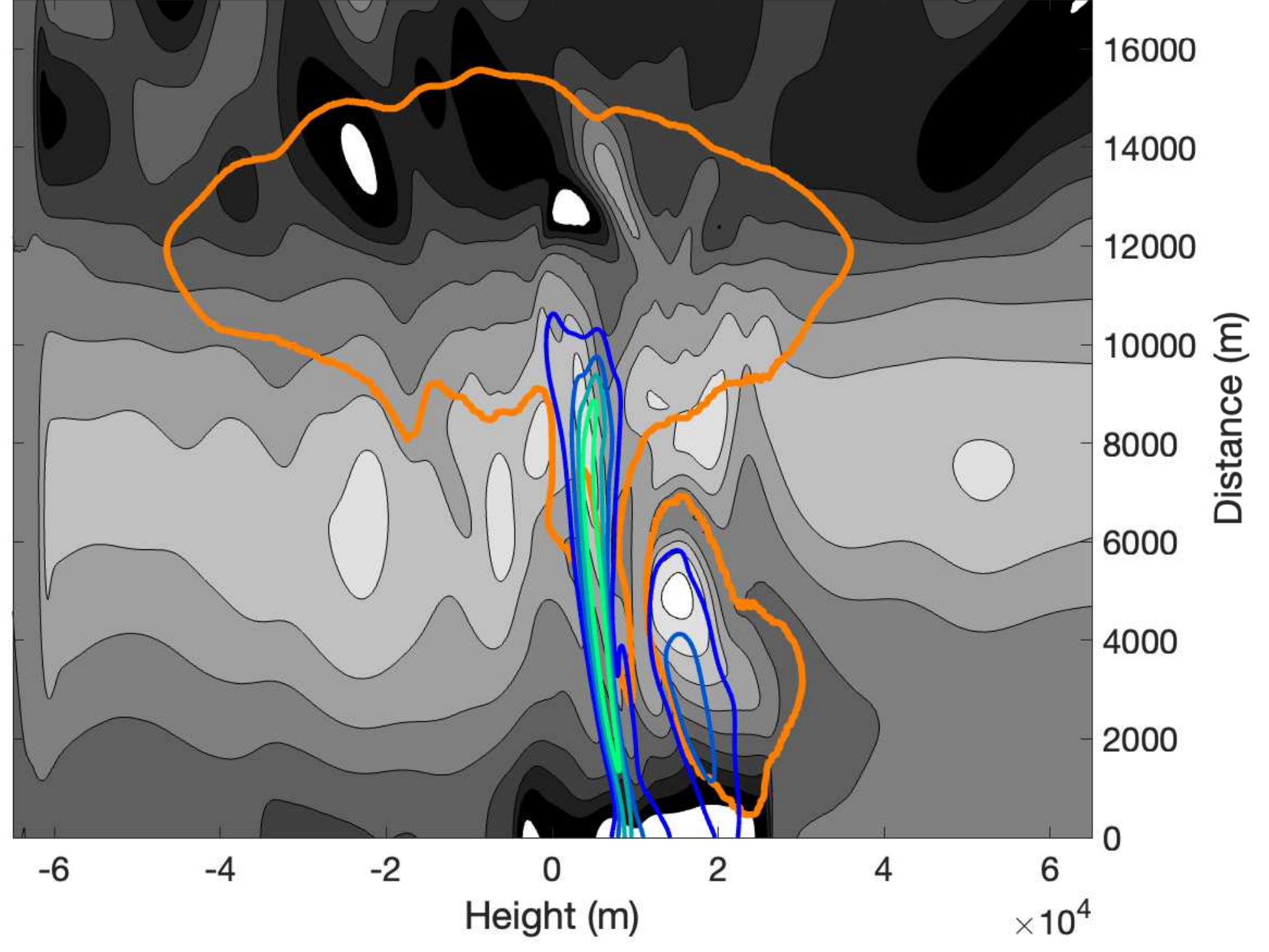} & \includegraphics[width=0.47\textwidth]{figures/100m_cloud_rain_9000s.pdf} \\
           
            \includegraphics[width=0.448\textwidth]{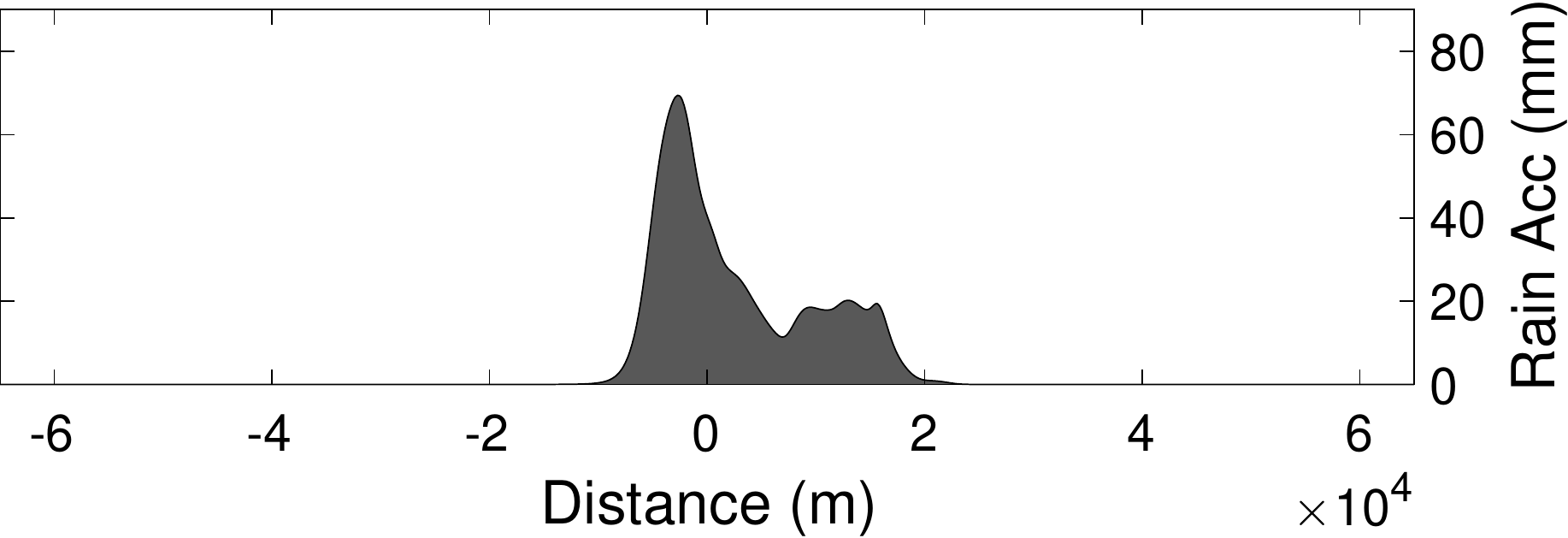} &  \includegraphics[width=0.448\textwidth]{figures/rain_accum_100m_9000s.pdf}
        \end{tabular}
    \caption{Storm at $t=9000~$s computed with the CG method and meshes $\Delta x = 250$~m (top-left), $\Delta x = 200$~m (top-right), $\Delta x= 150$~m (bottom-left), and $\Delta x = 100$~m (bottom-right).
    The thick orange contour line ($q_c= 10^{-5}~{\rm kgkg^{-1}}$) represents the outline of the cloud. The white and gray contours represent the perturbation potential temperature and the blue and green contours represent $q_r$. The bottom portions of each panel show the rain accumulated at the surface as a function of horizontal distance from the point $x=0$~m.}
    \label{storm_res_CG}
\end{figure}

\begin{figure}
      \begin{tabular}{ll}
    \includegraphics[width=0.4\textwidth]{figures/pot_temp_colorbar.pdf} &
    \includegraphics[width=0.4\textwidth]{figures/cloud_scale_rain.pdf}
    \\
    \includegraphics[width=0.47\textwidth]{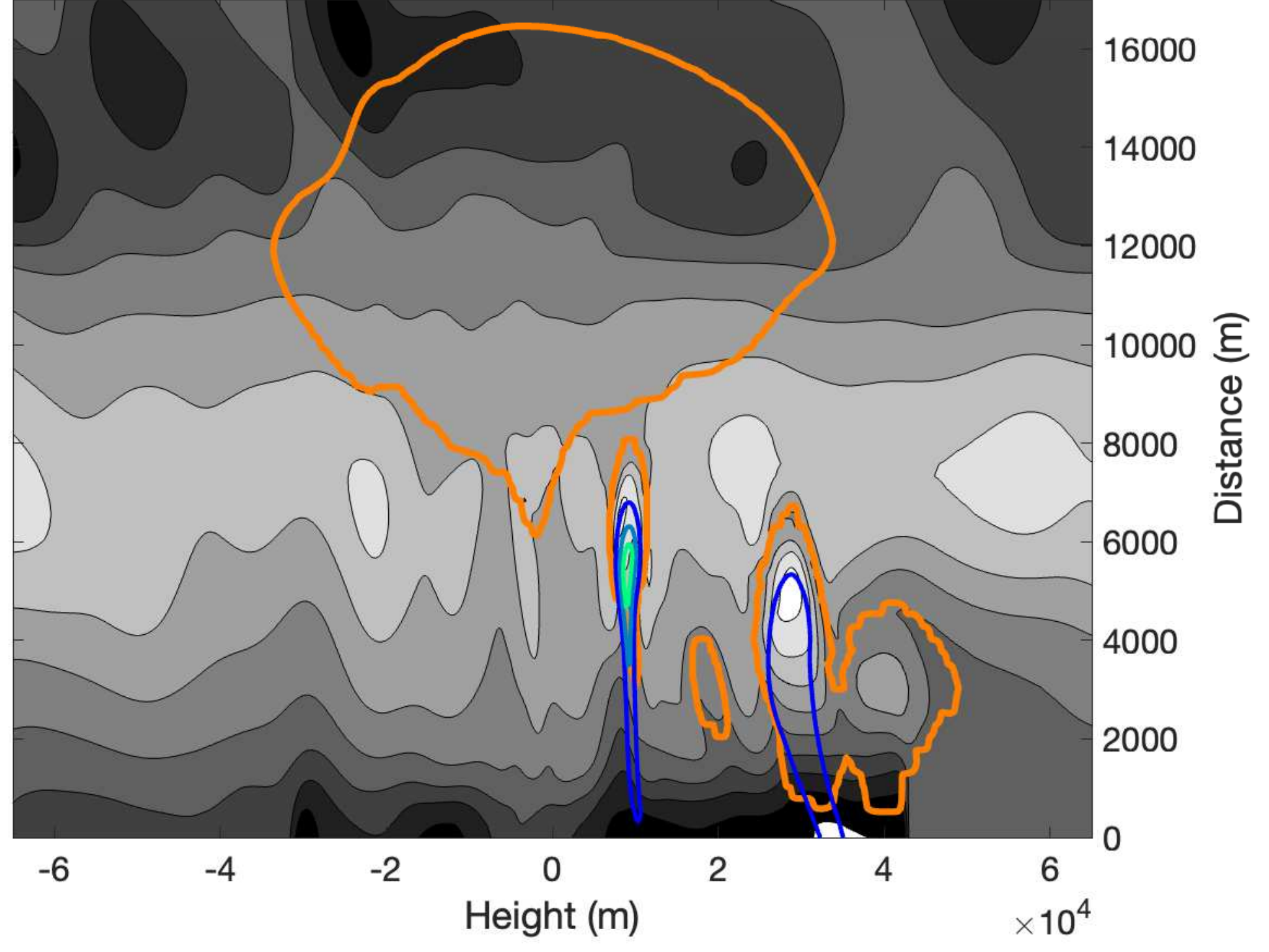} & \includegraphics[width=0.47\textwidth]{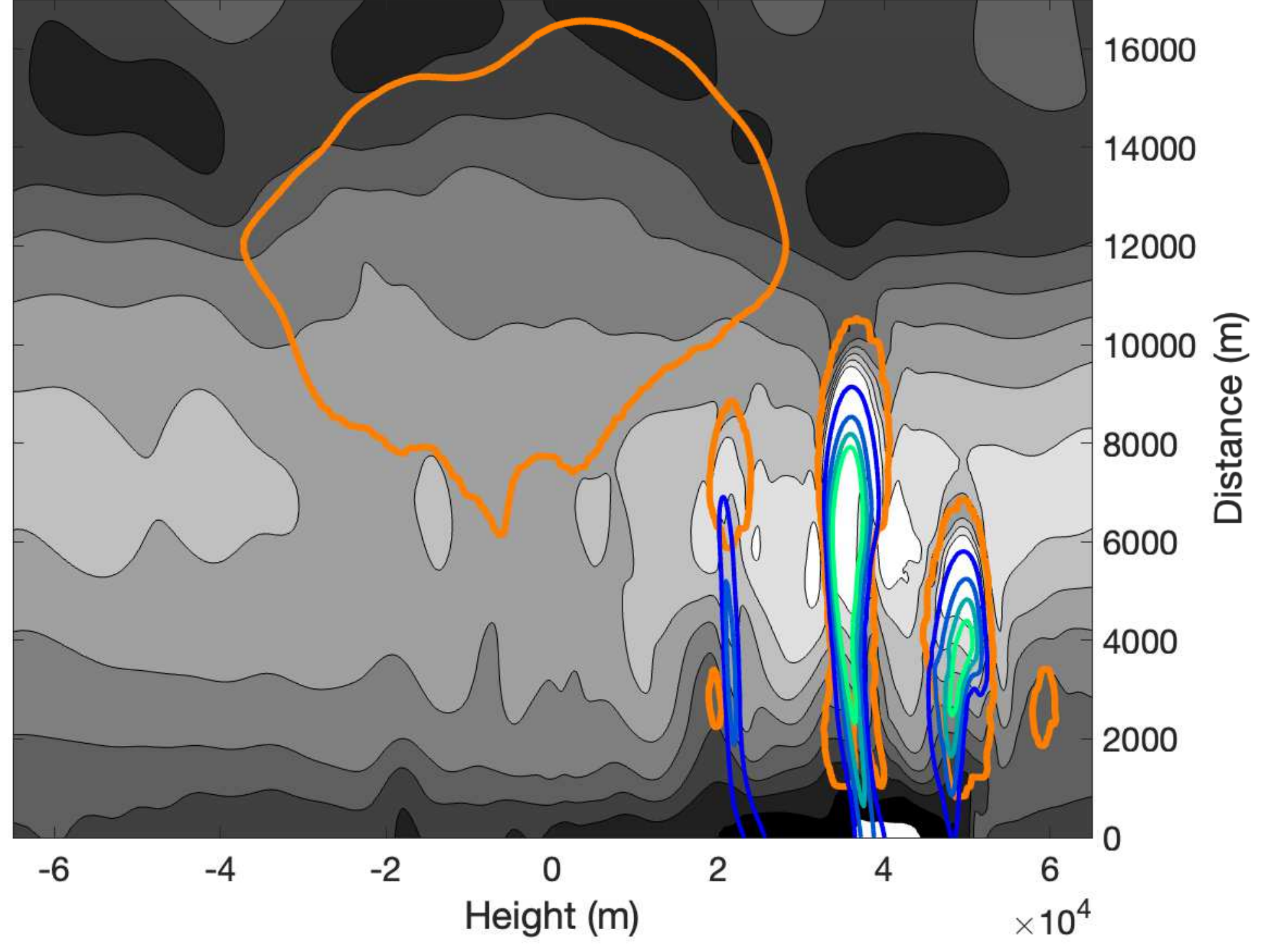} \\
           
            \includegraphics[width=0.448\textwidth]{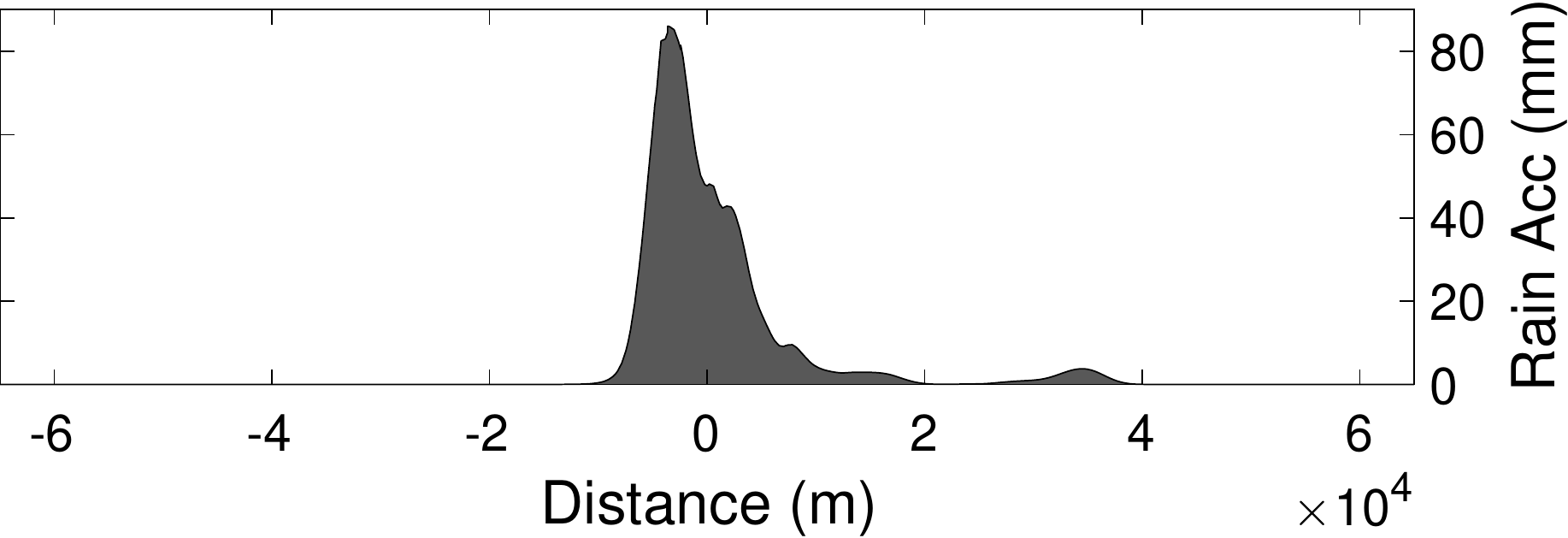} &  \includegraphics[width=0.448\textwidth]{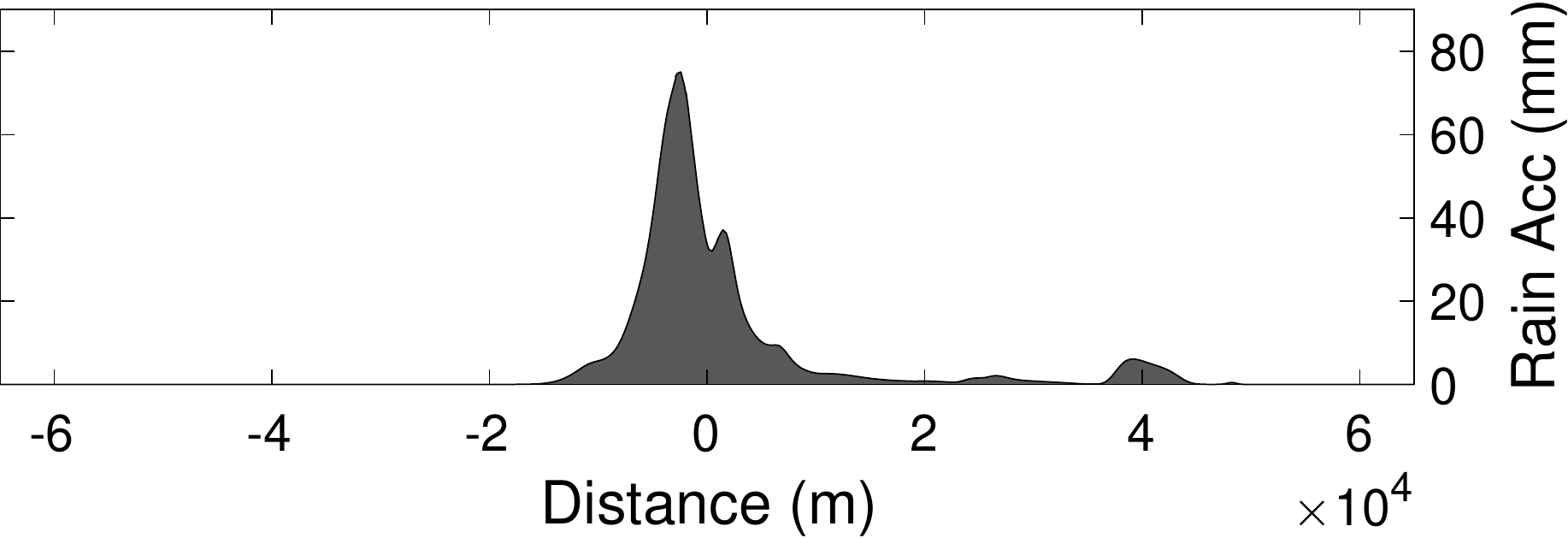}\\
            
            \includegraphics[width=0.47\textwidth]{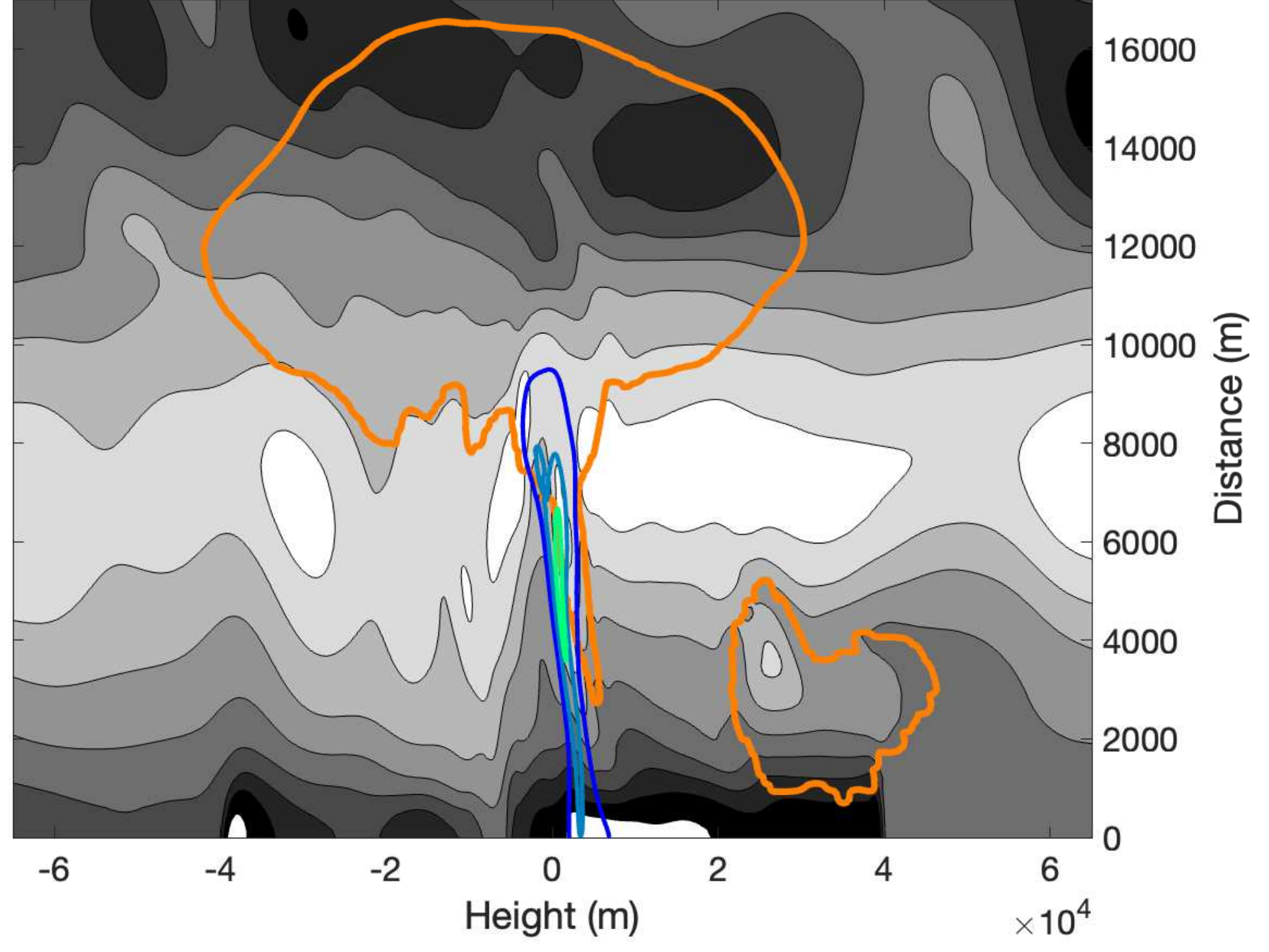} & \includegraphics[width=0.47\textwidth]{figures/100m_dg_cloud_rain_9000s.pdf} \\
           
            \includegraphics[width=0.448\textwidth]{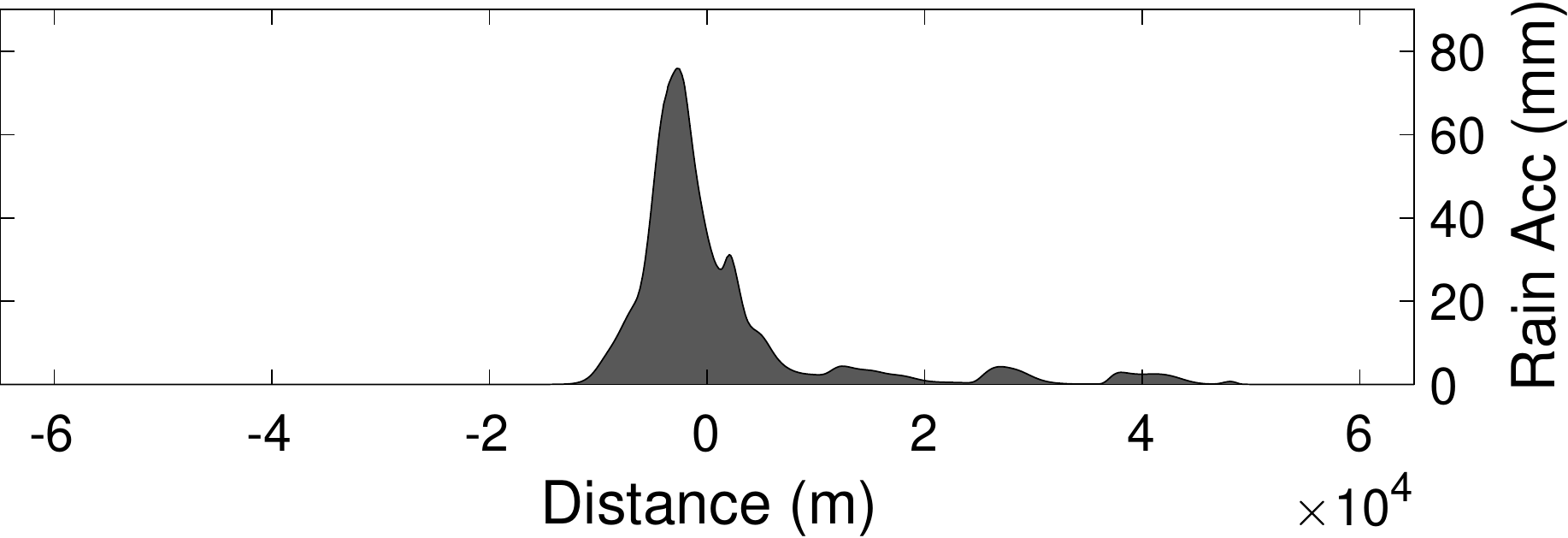} &  \includegraphics[width=0.448\textwidth]{figures/rain_accum__dg_100m_9000s.pdf}
        \end{tabular}
    \caption{Storm at $t=9000~$s computed with the DG method and meshes $\Delta x = 250$~m (top-left), $\Delta x = 200$~m (top-right), $\Delta x= 150$~m (bottom-left), and $\Delta x = 100$~m (bottom-right).
    The thick orange contour line ($q_c= 10^{-5}~{\rm kgkg^{-1}}$) represents the outline of the cloud. The white and gray contours represent the perturbation potential temperature and the blue and green contours represent $q_r$. The bottom portions of each panel show the rain accumulated at the surface as a function of horizontal distance from the point $x=0$~m.}
    \label{storm_res_DG}
\end{figure}

The results obtained with the $\Delta x = 250, 200, 150, 100$~m meshes at $t = 9000$ s are compared in Fig.~\ref{storm_res_CG} for the CG approximation and in 
Fig.~\ref{storm_res_DG} for the DG approximation.
In Fig.~\ref{storm_res_CG}, we observe the same cloud structure (anvil extent, downwind tilt of the convective tower) and similar profiles of perturbation potential temperature
for all the meshes under consideration. However, the spatial distributions of the rainfall accumulated at the ground show some differences: the simulations with
resolutions $\Delta x =250$~m and $\Delta x=200$~m have smaller peaks of rain accumulation near the domain center than the simulations with
$\Delta x =150$~m and $\Delta x=100$~m.
The simulations with the $\Delta x=290, 500,750$~m meshes (not shown for brevity) give even more intense rainfall than 
the $\Delta x =250$~m and $\Delta x=200$~m simulations. A similar observation on rain accumulation and mesh resolution for this benchmark can be found in \cite{weismanEtAl1997,gabersekGiraldoDoyle2012}, where it is shown that higher resolutions are correlated with faster storm development, weaker storm circulation and less overall precipitation over the length of the simulation. 
The DG simulations also show similar tilt in the convective tower, similar anvil extents and similar profiles of perturbation potential temperature at $t = 9000$ s for all the meshes; see Fig.~\ref{storm_res_DG}. Concerning the rain accumulation, the DG simulation with the $\Delta x =250$~m mesh gives a very large primary and secondary peak near the center of the domain. The amount of rain falling at the domain center decreases with increasing resolution. Indeed, the $\Delta x =200, 150$~m simulations give a smaller amount of accumulated rain in the domain center and slightly larger peaks downwind and away from the center, reflecting the availability of more moisture for the secondary convective tower. 
Once again, we observe a decrease in precipitation with increasing resolution 
as expected \cite{gabersekGiraldoDoyle2012,weismanEtAl1997,marrasEtAl2013a,marrasGiraldo2014}.

\begin{figure}[h!]
    \centering
    \includegraphics[width =0.5\textwidth]{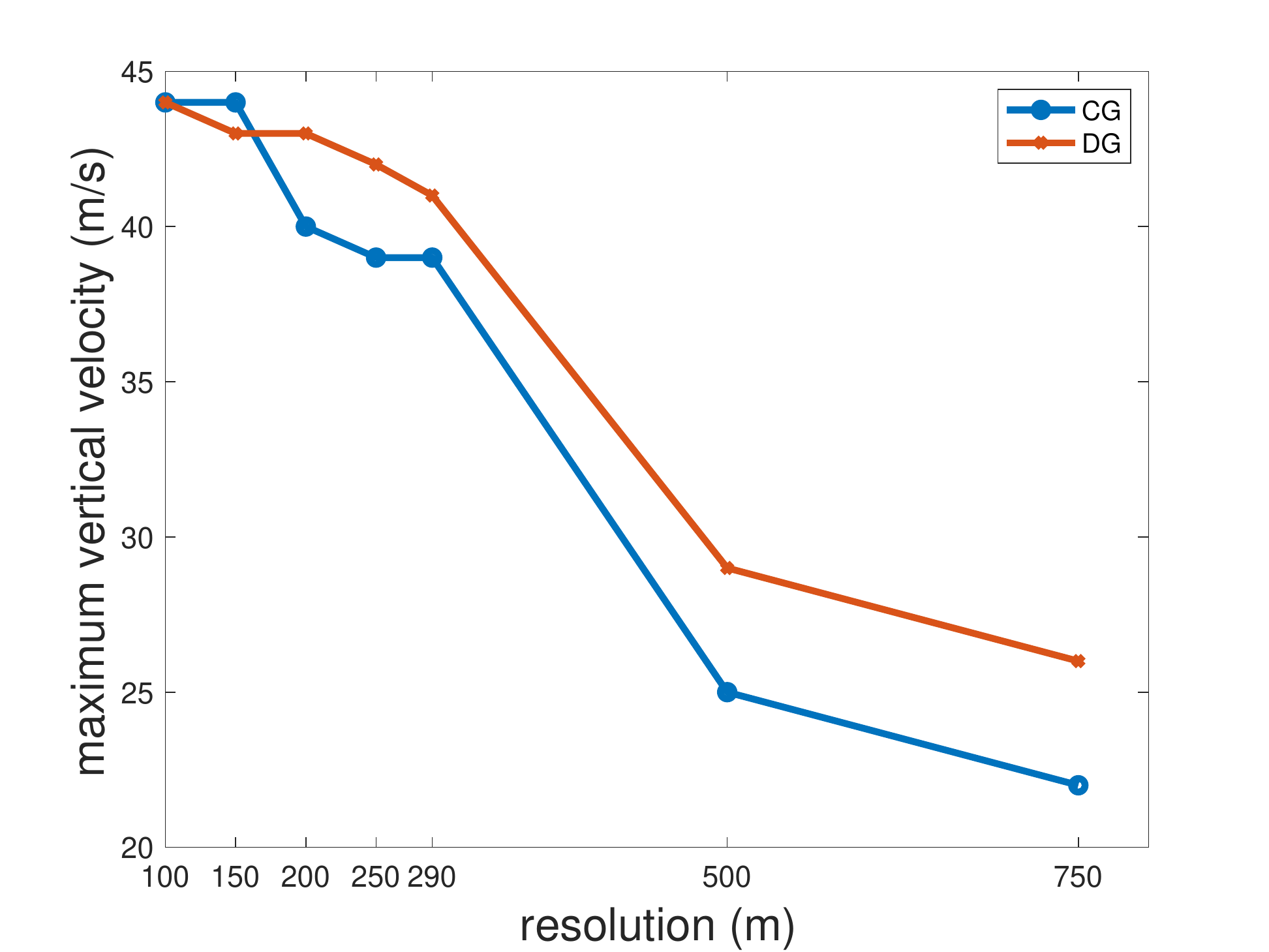}
    \caption{Maximum vertical velocity obtained over the course of the CG and DG simulations as a function of the resolution.}
    \label{fig:wmax}
\end{figure}

We conclude by reporting the maximum vertical velocity obtained over the course of the CG and DG simulation as a function of the resolution
in Fig. \ref{fig:wmax}. We see that for $\Delta x \geq 290~$m the maximum vertical velocity for both DG and CG simulations lies between $20~ {\rm ms^{-1}}$ and $30~ {\rm ms^{-1}}$, as in \cite{Bryanetal2006,WeismanRotunno2004,gabersekGiraldoDoyle2012}. Increasing the resolution yields an increase in the maximum velocity, 
as shown in \cite{gabersekGiraldoDoyle2012}. 
We note that the CG and DG simulations give similar values of the maximum vertical velocity
for a given mesh, with the values getting closer as the resolution increases.

The results in this section demonstrate that 
our algorithm successfully transports the rain downwards along the convective towers without the need for a vertically structured grid. 


\subsection{3D supercell}

In this section we test our algorithm for a fully three-dimensional supercell. The convective cell develops within a domain $\Omega =[150\times 100 \times 24]~ {\rm km^3}$. The storm is initiated by a thermal perturbation of the background state defined by \eqref{warmEqn}, with center $(x_c,y_c,z_c) = (75000,50000,2000)$~m and
\[
r = \sqrt[]{\frac{(x-x_{c})^{2}}{r_x^2}+\frac{(y -y_{c})}{r_y^2} + \frac{(z-z_{c})^{2}}{r_z^2}}, \quad \theta_c = 3~\text{K}, \quad r_c= 1,
\]
where:
\[r_x=r_y=10000~\text{m}, \quad r_z = 2000~\text{m}.\]

\begin{figure}[htb!]
    \centering
     \includegraphics[width = \textwidth]{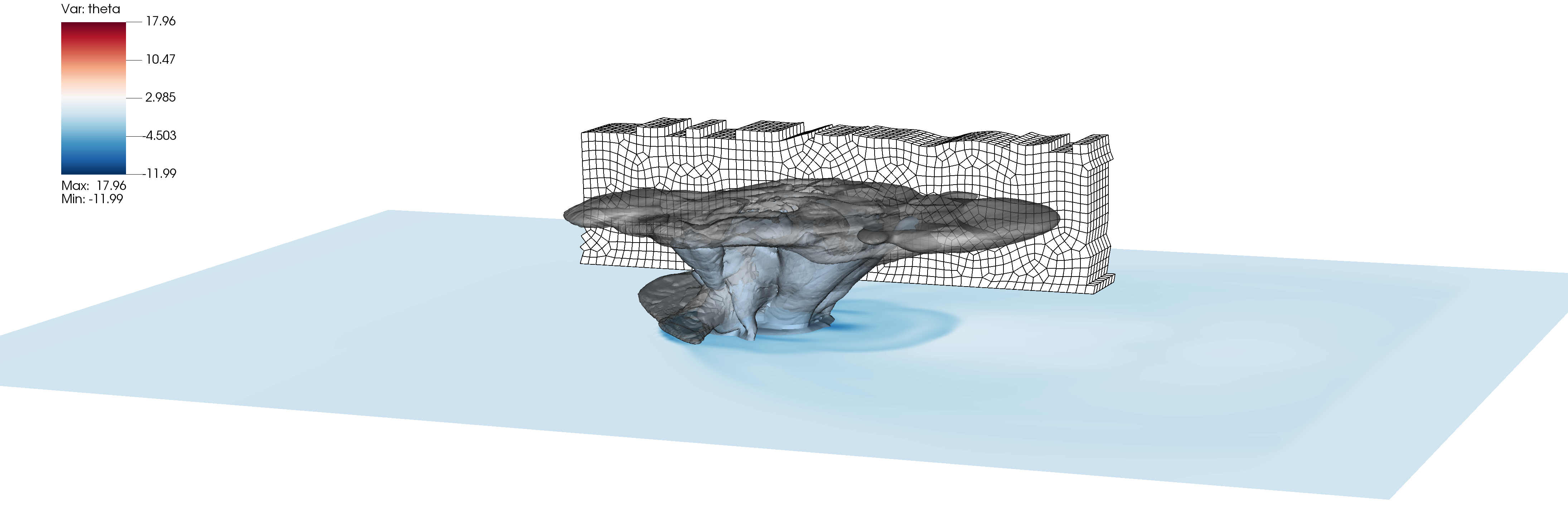}
    \caption{3D mature supercell at $t=7200~$s. The grey shading is the iso-surface $q_c=1e-5~{\rm kg/kg}$. The semi-transparent blue shading is the iso-surface $q_r=1e-4~{\rm kg/kg}$. The blue shading is the perturbation potential temperature (blue is negative) showing the cold pools due to rain evaporation. A small sample of the three-dimensional unstructured grid is shown in the background.}
    \label{3d_cloud_paraview}
\end{figure}

The domain is discretized using a grid of unstructured hexahedra of order 4 in all directions for an approximate effective resolution $\Delta {\bf x} \approx 250$~m. The grid is partially shown in Fig.~\ref{3d_cloud_paraview}.

We use periodic boundary conditions for the lateral boundaries, a free-slip boundary at the domain bottom and a Rayleigh sponge at the domain top. 
Like for the squall line test described above,
we use the ARK3 3D semi-implicit time integrator to advance the simulation in time and keep the acoustic Courant number $C\leq 1$. An artificial viscosity $\beta= 200$ (see Remark \ref{rem2} for the units) is used to provide stabilization. 
The wind shear in the $x$ direction is the same as the one used for the squall-line. The cloud begins to form at $t\approx 500~$s while rain forms and  starts to precipitate at $t\approx 900~$s. 

A 3D view of the fully developed storm at $t=7200$~s is shown in Fig.~\ref{3d_cloud_paraview}, along with a partial view of the three-dimensional grid. The semi-transparent blue shading is the iso-surface $q_r=1e-4~{\rm kg/kg}$. The blue shading is the perturbation potential temperature (blue is negative) showing the cold pools due to rain evaporation. All of the convective towers exhibit tilting due to wind-shear, with the parts closer to the ground experiencing a greater wind-shear and thus trailing the rest of the convective tower.
An anvil cloud is also observed near the top of the troposphere.

\begin{figure}[htb!]
    \centering
    \includegraphics[width = 0.49\textwidth]{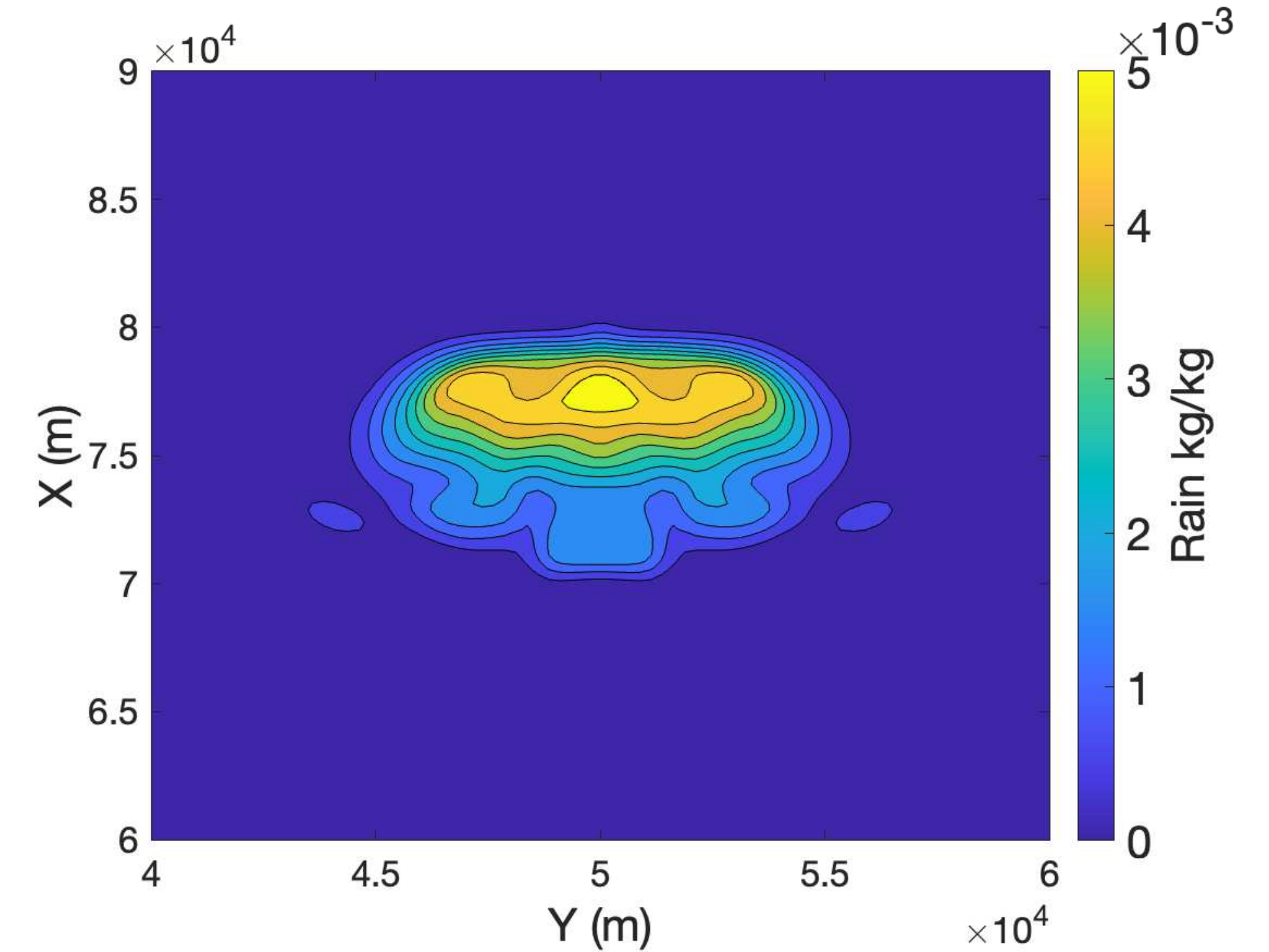}
    \includegraphics[width = 0.49\textwidth]{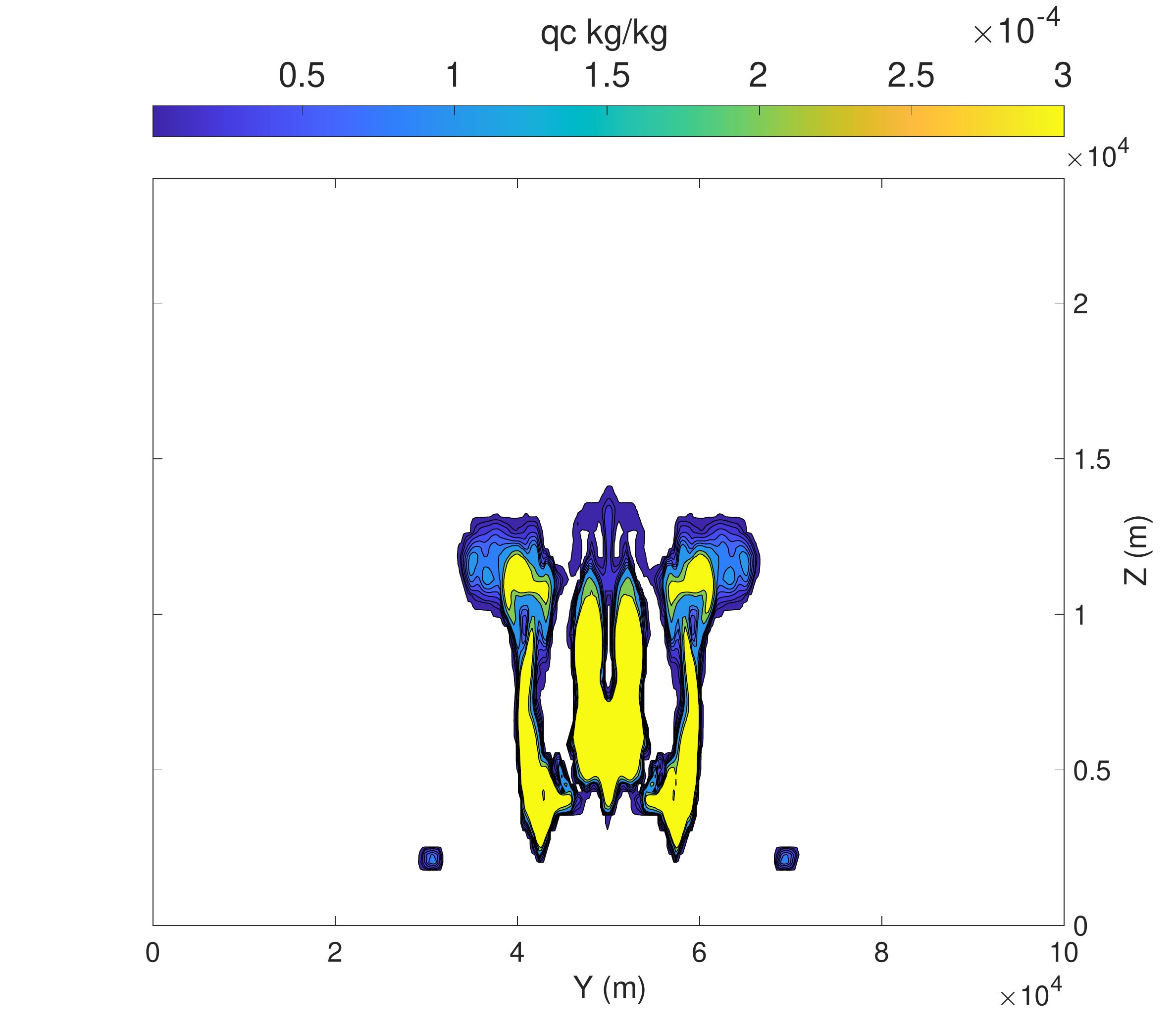}
    \caption{State of the storm at $t=7200$~s. Left: Horizontal cross-section of the instantaneous distribution of rain along the surface ($z=0$~m) at $t=7200~$s. Right: Vertical cross section taken at $x=75000$~m of the cloud fraction $t=7200$~s}
    \label{3d_cloud}
\end{figure}

Fig.~\ref{3d_cloud} shows the state of the storm at $t=7200$~s. The right side of the figure shows the existence of 3 distinct convective towers in the supercell. One in the center of the Y axis at $y=50000$~m and two columns symmetric about $y=50000$~m plane. 
The three towers merge into the anvil cloud near the tropopause. Fig.~\ref{3d_cloud} (left) shows the rain distribution at the ground at $t=7200$~s. The position of the rain concentration follows the location of the convective towers, falling below them. The largest amount of rain is present below the larger central tower as indicated by the maximum over $y=50000$~m. Additionally we can see the presence of some rain slightly separated from the main rain distribution which corresponds to the small low clouds that are shown symmetric to the $y=50000$~m plane in the right side of the figure.

The results presented in this section show that
the storm develops in a symmetrical manner and the rain falls correctly following the location of the convective towers, as is expected. This is accomplished without a column based grid.
This demonstrates that our algorithm successfully transports the rain downward along the convective towers without the need for a vertically structured grid also in three dimensions.


\section{Conclusions}\label{sec:conclusions}

We presented an algorithm to solve the transport equation of precipitating clouds and Kessler's microphysical processes on fully unstructured grids. The Euler equations of moist atmospheric flows (embedded with artificial diffusion for stabilization purposes) were discretized by $4^{th}$-order continuous and discontinuous spectral elements in space and advanced in time by a $3^{rd}$-order additive Runge-Kutta semi-implicit time integrator.
The results of these simulations are in very good agreement with results in the literature obtained using vertically structured meshes and column-based microphysics. This shows that the algorithm, while simple, does succeed in handling moisture with unstructured grids.

Coupled with the flexibility of the spectral element method, we believe that our algorithm could 
successfully resolve storms over steep terrain \cite{mariscoStechmann2020} using unstructured meshes with and without adaptive mesh refinement, without the need for a special physics grid on which to handle moisture. Work in this direction is recommended. While we presented results only for warm rain, extension to other moist precipitation processes is natural.
Probably the greatest advantage of fully unstructured atmospheric simulations is the fact that parallel load balancing decomposition can be done in any direction, which is of fundamental importance for efficient exascale simulations of high-resolution  weather and climate modeling.

\section{Data Availability Statement}
All data presented in the paper and the source code with the unstructured algorithm are available on a public github repository through Zenodo via this DOI \url{https://doi.org/10.5281/zenodo.6787870}, with the GNU General Public License v3.0.

\section*{Author contributions}
{\bf Yassine Tissaoui}: Methodology, Software, Validation, Formal Analysis, Investigation, Visualization, Writing. \textbf{Simone Marras, PI}: Conceptualization, Methodology, Software,  Writing, Review, Editing, Supervision. \textbf{Annalisa Quaini}: Writing, Review, Editing. \textbf{Felipe A. V. de Braganca}: Software. \textbf{Francis X. Giraldo}. Software, Writing, Review, Editing. 

\section*{Acknowledgments}
The authors are grateful to Dr. James F. Kelly from the U.S. Naval Research Laboratory (Washington, DC) for his feedback on the manuscript. Yassine Tissaoui and Simone Marras acknowledge the partial support by the National Science Foundation through grant PD-2121367. 
Annalisa Quaini acknowledges partial support by the National Science Foundation through grant DMS-1953535 and support from the Radcliffe Institute for Advanced Study at Harvard University where she has been the 2021-2022 William and Flora Hewlett Foundation Fellow.
Francis Giraldo gratefully acknowledges the support of ONR under grant \# N0001419WX00721. Francis Giraldo and Felipe Alves gratefully acknowledge support from the Defense Sciences Office of the Defense Applied Research and Projects Agency (DARPA DSO) through the Space Environment Exploitation (SEE) program. This work was performed
when Felipe Alves held a National Academy of Sciences’ National Research Council (NRC) Fellowship at the Naval Postgraduate School.
Yassine Tissaoui and Simone Marras gratefully acknowledge the Extreme Science and Engineering Discovery Environment (XSEDE) for providing core hours on Bridges-2 with allocation TG-EES210027.


%
%


\bibliographystyle{ieeetr} 
\bibliography{Arxiv-formatted}
\newpage
\section*{Appendix}
\begin{table}[h!]
\centering
\caption{Squall line sounding}
\label{tab:sound}
\begin{tabular}{cccccc}
  \hline
  $z$ (m) & $\theta$ (K) & $q_v$ (g/kg) & $u$ (m/s) & $v$ (m/s) & $p$ (Pa) \\
  \hline
  0.0 &  303.025079 & 14.000 & 12.0 & 0.0 & 100000.0 \\
480.0 &  303.337272 & 14.000 & 9.696000 & 0.0 & 94697.28 \\
960.0 &  304.402985 & 14.000 & 7.392000 & 0.0 & 89609.81 \\
1440.0 & 305.397187 & 12.796 & 5.088000 & 0.0 & 84736.79 \\
1920.0 & 306.306214 & 10.556 & 2.784000 & 0.0 & 80070.30 \\
2400.0 & 307.365269 &  8.678 & 0.540000 & 0.0 & 75604.36 \\
2880.0 & 308.550318 &  7.104 & 0.0 & 0.0 & 71334.51 \\
3360.0 & 309.845257 &  5.788 & 0.0 & 0.0 & 67255.79 \\
3840.0 & 311.235047 &  4.691 & 0.0 & 0.0 & 63362.95 \\
4320.0 & 312.708238 &  3.777 & 0.0 & 0.0 & 59650.49 \\
4800.0 & 314.255743 &  3.020 & 0.0 & 0.0 & 56112.80 \\
5280.0 & 315.869985 &  2.396 & 0.0 & 0.0 & 52744.15 \\
5760.0 & 317.544512 &  1.885 & 0.0 & 0.0 & 49538.82 \\
6240.0 & 319.273784 &  1.469 & 0.0 & 0.0 & 46491.09 \\
6720.0 & 321.052868 &  1.134 & 0.0 & 0.0 & 43595.27 \\
7200.0 & 322.877588 &  0.866 & 0.0 & 0.0 & 40845.73 \\
7680.0 & 324.744235 &  0.653 & 0.0 & 0.0 & 38236.93 \\
8160.0 & 326.649534 &  0.487 & 0.0 & 0.0 & 35763.41 \\
8640.0 & 328.590559 &  0.357 & 0.0 & 0.0 & 33419.84 \\
9120.0 & 330.565013 &  0.259 & 0.0 & 0.0 & 31200.99 \\
9600.0 & 332.571020 &  0.184 & 0.0 & 0.0 & 29101.75 \\
10080.0 & 334.606102 & 0.129 & 0.0 & 0.0 & 27117.17 \\
10560.0 & 336.668475 & 0.088 & 0.0 & 0.0 & 25242.39 \\
11520.0 & 340.869535 & 0.038 & 0.0 & 0.0 & 21803.59 \\
12000.0 & 343.712008 & 0.026 & 0.0 & 0.0 & 20232.15 \\
12480.0 & 350.647306 & 0.026 & 0.0 & 0.0 & 18763.71 \\
12960.0 & 358.453724 & 0.029 & 0.0 & 0.0 & 17401.15 \\
13440.0 & 366.433620 & 0.031 & 0.0 & 0.0 & 16138.11 \\
13920.0 & 374.591035 & 0.034 & 0.0 & 0.0 & 14967.29 \\
14400.0 & 382.929618 & 0.037 & 0.0 & 0.0 & 13881.93 \\
15360.0 & 400.170355 & 0.044 & 0.0 & 0.0 & 11942.99 \\
15840.0 & 409.081924 & 0.049 & 0.0 & 0.0 & 11078.24 \\
16320.0 & 418.191751 & 0.053 & 0.0 & 0.0 & 10276.53 \\
16800.0 & 427.504224 & 0.058 & 0.0 & 0.0 & 9533.23 \\
17280.0 & 437.023716 & 0.063 & 0.0 & 0.0 & 8844.07 \\
17760.0 & 446.755038 & 0.069 & 0.0 & 0.0 & 8205.09 \\
18720.0 & 466.871821 & 0.083 & 0.0 & 0.0 & 7063.24 \\
19200.0 & 477.267160 & 0.091 & 0.0 & 0.0 & 6553.82 \\
19680.0 & 487.891998 & 0.094 & 0.0 & 0.0 & 6081.42 \\
20160.0 & 498.742611 & 0.094 & 0.0 & 0.0 & 5643.35 \\
20640.0 & 509.643457 & 0.094 & 0.0 & 0.0 & 5237.00 \\
21120.0 & 520.544304 & 0.094 & 0.0 & 0.0 & 4859.92 \\
21600.0 & 531.445151 & 0.094 & 0.0 & 0.0 & 4509.85 \\
22560.0 & 553.246845 & 0.094 & 0.0 & 0.0 & 3882.66 \\
23040.0 & 564.147692 & 0.094 & 0.0 & 0.0 & 3601.93 \\
23520.0 & 575.048539 & 0.094 & 0.0 & 0.0 & 3340.96 \\
24000.0 & 585.949386 & 0.094 & 0.0 & 0.0 & 3098.30 \\
\end{tabular}
\end{table}

%
%
%
%
%
\end{document}